\documentclass{emulateapj}

\usepackage{amsmath}

\newcommand{\cm}{\ensuremath{\mbox{cm}}}
\newcommand{\erg}{\ensuremath{\mbox{erg}}}
\newcommand{\ev}{\ensuremath{\mbox{eV}}}
\newcommand{\K}{\ensuremath{\mbox{K}}}
\newcommand{\kev}{\ensuremath{\mbox{keV}}}
\newcommand{\km}{\ensuremath{\mbox{km}}}

 \newcommand{\pc}{\ensuremath{\mbox{pc}}}
\newcommand{\ph}{\ensuremath{\mbox{photons}}}
\newcommand{\s}{\ensuremath{\mbox{s}}}
\newcommand{\sr}{\ensuremath{\mbox{sr}}}
\newcommand{\yr}{\ensuremath{\mbox{yr}}}

\newcommand{\cmsq}{\ensuremath{\cm^2}}
\newcommand{\pcc}{\ensuremath{\cm^{-3}}}
\newcommand{\pcmsq}{\ensuremath{\cm^{-2}}}
\newcommand{\pkev}{\ensuremath{\kev^{-1}}}
\newcommand{\ps}{\ensuremath{\s^{-1}}}
\newcommand{\psr}{\ensuremath{\sr^{-1}}}
\newcommand{\emismeas}{\ensuremath{\cm^{-6}} \pc}
\newcommand{\flux}{\erg\ \pcmsq\ \ps}
\newcommand{\kmps}{\km\ \ps}
\newcommand{\lineunit}{\ph\ \pcmsq\ \ps\ \psr}
\newcommand{\lineunitabbr}{\ensuremath{\mathrm{ph}}\ \pcmsq\ \ps\ \psr}
\newcommand{\presalt}{\pcc\ \K}

\newcommand{\chisq}{\ensuremath{\chi^2}}
\newcommand{\citepossessive}[1]{\citeauthor{#1}'s \citeyearpar{#1}}
\newcommand{\citepsq}[1]{[\citealp{#1}]}
\newcommand{\citetsq}[1]{\citeauthor{#1} [\citeyear{#1}]}
\newcommand{\cs}{\ensuremath{c_\mathrm{s}}}
\newcommand{\dd}{\ensuremath{\mathrm{d}}}
\newcommand{\dl}{\ensuremath{\mathrm{d}l}}
\newcommand{\ftest}{\textit{F}-test}
\newcommand{\Kalpha}{K$\alpha$}
\newcommand{\logT}{\ensuremath{\log(T/\mathrm{K})}}
\newcommand{\Lyalpha}{Ly$\alpha$}
\newcommand{\mekal}{\textsc{MeKaL}}
\newcommand{\Ne}{\ensuremath{n_{\mathrm{e}}}}
\newcommand{\NH}{\ensuremath{N_{\mathrm{H}}}}
\newcommand{\nH}{\ensuremath{n_{\mathrm{H}}}}
\newcommand{\nHe}{\ensuremath{n_{\mathrm{He}}}}
\newcommand{\nO}{\ensuremath{n_{\mathrm{O}}}}
\newcommand{\rchisq}{\ensuremath{\chi^2_\nu}}
\newcommand{\TLB}{\ensuremath{T_\mathrm{LB}}}
\newcommand{\logTLB}{\ensuremath{\log(\TLB/\mathrm{K})}}

\newcommand{\asca}{\textit{ASCA}}
\newcommand{\chandra}{\textit{Chandra}}
\newcommand{\fuse}{\textit{FUSE}}
\newcommand{\iras}{\textit{IRAS}}
\newcommand{\rosat}{\textit{ROSAT}}
\newcommand{\suzaku}{\textit{Suzaku}}
\newcommand{\xmm}{\textit{XMM-Newton}}

\newcommand{\HII}{H~\textsc{ii}}
\newcommand{\NaI}{Na~\textsc{i}}
\newcommand{\OI}{O~\textsc{i}}
\newcommand{\OII}{O~\textsc{ii}}
\newcommand{\OV}{O~\textsc{v}}
\newcommand{\OVI}{O~\textsc{vi}}
\newcommand{\OVII}{O~\textsc{vii}}
\newcommand{\OVIII}{O~\textsc{viii}}
\newcommand{\OIX}{O~\textsc{ix}}

\newcommand{\Roviii}{\ensuremath{R_\mathrm{O\,VIII}}}
\newcommand{\Loviii}{\ensuremath{L_\mathrm{O\,VIII}}}
\newcommand{\Boviii}{\ensuremath{B_\mathrm{O\,VIII}}}
\newcommand{\taueon}{\ensuremath{\tau^\mathrm{on}_\mathrm{O\,VIII}}}
\newcommand{\taueoff}{\ensuremath{\tau^\mathrm{off}_\mathrm{O\,VIII}}}

\newcommand{\Rovii}{\ensuremath{R_\mathrm{O\,VII}}}
\newcommand{\Lovii}{\ensuremath{L_\mathrm{O\,VII}}}
\newcommand{\Bovii}{\ensuremath{B_\mathrm{O\,VII}}}
\newcommand{\tauson}{\ensuremath{\tau^\mathrm{on}_\mathrm{O\,VII}}}
\newcommand{\tausoff}{\ensuremath{\tau^\mathrm{off}_\mathrm{O\,VII}}}

\newcommand{\Ioviii}{\ensuremath{I_\mathrm{O\,VIII}}}
\newcommand{\Iovii}{\ensuremath{I_\mathrm{O\,VII}}}
\newcommand{\Iovi}{\ensuremath{I_\mathrm{O\,VI}}}

\newcommand{\noviii}{\ensuremath{n_\mathrm{O\,VIII}}}
\newcommand{\novi}{\ensuremath{n_\mathrm{O\,VI}}}

\shorttitle{AN \textit{XMM-NEWTON} OBSERVATION OF THE LOCAL BUBBLE}
\shortauthors{HENLEY, SHELTON, AND KUNTZ}

\begin{document}

\title{An \textit{XMM-Newton} Observation of the Local Bubble using a Shadowing Filament in the Southern Galactic Hemisphere}
\author{David B. Henley and Robin L. Shelton}
\affil{Department of Physics and Astronomy, University of Georgia, Athens, GA 30602}
\and
\author{K. D. Kuntz}
\affil{Henry A. Rowland Department of Physics and Astronomy, Johns Hopkins University, Baltimore, MD 21218}
\affil{Exploration of the Universe Division, NASA Goddard Space Flight Center, Code 662, Greenbelt, MD 20771}

\email{dbh@physast.uga.edu}

\begin{abstract}

We present an analysis of the X-ray spectrum of the Local Bubble, obtained by simultaneously analyzing spectra from two \xmm\
pointings on and off an absorbing filament in the Southern galactic hemisphere ($b \approx -45\degr$). We use the difference
in the Galactic column density in these two directions to deduce the contributions of the unabsorbed foreground emission due
to the Local Bubble, and the absorbed emission from the Galactic halo and the extragalactic background. We find the Local Bubble
emission is consistent with emission from a plasma in collisional ionization equilibrium with a temperature
$\logTLB = 6.06^{+0.02}_{-0.04}$ and an emission measure $\int \Ne^2 \dl = 0.018$ \emismeas.
Our measured temperature is in good agreement with values obtained from \rosat\ All-Sky Survey data, but is lower than that measured by
other recent \xmm\ observations of the Local Bubble, which find $\logTLB \approx 6.2$ (although for some of these observations it is
possible that the foreground emission is contaminated by non-Local Bubble emission from Loop I). The higher temperature observed
towards other directions is inconsistent with our data, when combined with a \fuse\ measurement of the Galactic halo \OVI\ intensity.
This therefore suggests that the Local Bubble is thermally anisotropic.

Our data are unable to rule out a non-equilibrium model in which the plasma is underionized. However, an overionized
recombining plasma model, while observationally acceptable for certain densities and temperatures, generally gives an implausibly
young age for the Local Bubble ($\la 6 \times 10^5$ \yr).

\end{abstract}

\keywords{Galaxy: general---Galaxy: halo---ISM: general---ISM: individual (Local Bubble)---X-rays: ISM}


\section{INTRODUCTION}
\label{sec:Introduction}

The Local Bubble (LB) is a region of X-ray--emitting gas of $\sim$100 \pc\ extent in which the Solar System resides.
Measurements of the interstellar \NaI\ absorption towards 456 nearby stars reveal that the Local Bubble resides in a cavity
in the interstellar medium (ISM; \citealp{sfeir99}). The idea
of the Local Bubble originated in the late 1970s, in order to explain the observed anticorrelation between the intensity of the soft X-ray
background (SXRB) and the Galactic hydrogen column density \NH\ \citep{bowyer68,sanders77}. The data are inconsistent with the
anticorrelation being due to absorption, as they require an interstellar absorption cross-section that is one-third of its
expected value \citep{bowyer68}, and also different energy bands have the same dependence on \NH\ (\citealp{sanders77}; \citealp{juda91}).
Instead, a so-called ``displacement'' model was proposed, in which the hot X-ray--emitting plasma (the Local Bubble) is in
the foreground and displaces the cool gas \citep{sanders77, tanaka77}.
In directions of higher X-ray intensity the Local Bubble is thought to be of greater extent, and so there is less cool gas (and hence
lower \NH) in those directions.

Numerous models have been proposed for the formation of the Local Bubble (for reviews, see \citealt{breitschwerdt98}; \citealt{cox98}; \citealt{breitschwerdt04}).
These models may essentially be divided into two classes. In one class of model, the Local Bubble was carved out of the ambient ISM by a supernova or series of
supernovae (e.g.\ \citealp{cox82}; \citealp{innes84}; \citealp{smith01b}; \citealp{maizapellaniz01}; \citealp{breitschwerdt06}). The hot gas thus produced
gives rise to the observed X-rays.  If the last supernova was recent enough, one would expect the ions to be underionized. However, if the Local Bubble is old
enough to have begun contracting, the ions will be overionized and recombining \citep{smith01b}.
In the second class of model, a series of supernovae in a dense cloud formed a hot bubble, which burst out of the cloud into the less dense
surroundings and underwent rapid adiabatic cooling (\citealp{breitschwerdt94}; \citealp{breitschwerdt96a}; \citealp{breitschwerdt96b}; \citealp{breitschwerdt01}). In this case
the X-ray emission is due to the delayed recombination of the overionized ions.
X-ray spectroscopy of the Local Bubble emission is essential for distinguishing between the various models, as it enables us to determine
the physical properties of the X-ray--emitting gas.

Originally, all the soft X-ray flux was attributed to the Local Bubble, which made determining the Local Bubble X-ray spectrum relatively simple. However, the discovery
of shadows in the SXRB with \rosat\ (\citealp{snowden91}; \citealp{burrows91}) showed that $\sim$50\%\ of the SXRB in the 1/4-\kev\ band originated from beyond the Local Bubble,
either in the Galactic halo or from an extragalactic background. Hence, in order to determine the Local Bubble spectrum, one must first disentangle the contributions
of the Local Bubble and the background. This has been done by modeling \rosat\ All-sky Survey data (RASS; \citealp{snowden97}) with an unabsorbed foreground component
(due to the Local Bubble) and absorbed components for the Galactic halo and the extragalactic background, using the 100-\micron\ 
\textit{Infrared Astronomical Satellite} (\iras) maps of \citet{schlegel98} as a measure of \NH. In this way the Local Bubble emission has been mapped out, and
its temperature estimated to be $\TLB \sim 10^6$ \K\ assuming collisional ionization equilibrium (\citealp{snowden98,snowden00}; \citealp{kuntz00}).

It should be noted, however, that the RASS data are presented in just six energy bands between $\sim$0.1 and $\sim$2 \kev, several of which overlap.
The data therefore have fairly poor spectral resolution. The Local Bubble temperature is inferred from the R1-to-R2 band intensity ratio of the local emission
component, as there is very little Local Bubble emission in the higher-energy bands: the \textit{upper limit} on the \rosat\ R45 intensity due to the Local Bubble is
a $\mbox{few} \times 10^5$ counts \ps\ arcmin$^{-2}$ (\citealp{snowden93}; \citealp{kuntz97}). However, this observational fact may be used to rule out
significantly higher temperatures. It should also be noted that this inferred temperature is somewhat dependent upon the plasma emission model used
(e.g.\ \citealp{kuntz00}).

The large mirror collecting area and CCD detectors onboard \xmm\ enable us to obtain high signal-to-noise spectra of the SXRB with greater
spectral resolution than \rosat. In particular, \xmm\ enables us to detect line emission in the SXRB spectrum, notably emission from \OVII\ and \OVIII\
at $\sim$0.55 and $\sim$0.65 \kev. If we can determine how much of this oxygen emission is due to the Local Bubble, this will enable us to place stronger constraints on \TLB.

\begin{figure}
\plotone{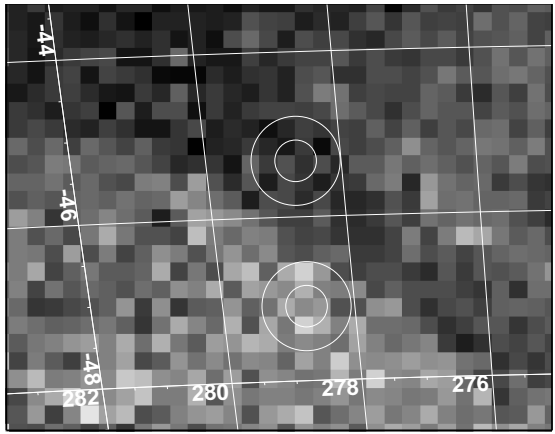}
\caption{\rosat\ All-Sky Survey R1$+$R2 image centered on $l = 279\degr$, $b = -46\degr$, showing the absorbing filament used
for our observations (data from \citealp{snowden97}). The circles show our on-filament (upper) and off-filament (lower)
pointing directions. The smaller circles ($\mathrm{radius} = 14\arcmin$) show the approximate areas from which our \xmm\
spectra were extracted, while the larger circles ($\mathrm{radius} = 30\arcmin$) show the areas from which our \rosat\ spectra
were extracted.\label{fig:FilamentImage}}
\end{figure}

We use a shadowing technique to determine the spectrum of the Local Bubble emission. We have analyzed spectra obtained from \xmm\ pointings on and off an absorbing
filament in the southern Galactic hemisphere. The absorbing filament appears as a shadow in the SXRB, as shown in Figure~\ref{fig:FilamentImage},
which also shows our \xmm\ pointing directions. \citet{penprase98} have estimated the distance of the filament to be $230 \pm 30$ \pc. Maps of the extent
of the Local Bubble, obtained from RASS data \citep{snowden98} and \NaI\ absorption data \citep{sfeir99,lallement03}, indicate that the boundary of the Local Bubble is $\sim$100 \pc\
away in this direction. Thus, the filament is between the Local Bubble and the Galactic halo. We fit spectral models simultaneously to the on- and off-filament spectra, using
the difference in the absorbing column between the two pointing directions
($\NH = 9.6 \times 10^{20}$ \pcmsq\ versus $1.9 \times 10^{20}$ \pcmsq; see \S\ref{subsec:ModelDescription}) to deduce the
contributions of the foreground (unabsorbed) and background (absorbed) model components to the
observed spectra.

Our observations and the data reduction are described in \S\ref{sec:Observations}.
The spectral models used to fit to the data are described in \S\ref{sec:Modeling}, and the fit results are presented
in \S\ref{sec:Results}. We discuss our results in \S\ref{sec:Discussion}. In particular, we compare our results with
the results of RASS analyses in \S\ref{subsec:ROSAT}, and we compare our results' prediction of the Galactic halo \OVI\
intensity with the observed value in \S\ref{subsec:Halo}. In \S\ref{subsec:OtherObs} we compare our results with those
of other \xmm\ and \chandra\ observations of the Local Bubble. In \S\ref{subsec:SpectralModel} we discuss our choice of plasma
emission code used in the analysis, and consider the effect this has on our results. Finally in this section we discuss
non-equilibrium models of the Local Bubble: an underionized model in \S\ref{subsec:NEI}, and a recombining model in \S\ref{subsec:Recombining}.
We conclude with a summary in \S\ref{sec:Summary}. Throughout this paper we quote $1\sigma$ errors.

\section{OBSERVATIONS AND DATA REDUCTION}
\label{sec:Observations}

The on- and off-filament \xmm\ observations were both carried out on 2002 May 3. The details of
the observations are presented in Table~\ref{tab:Observations}.

\begin{deluxetable*}{cccccccc}
\tablewidth{0pt}
\tablecaption{Details of the \textit{XMM-NEWTON} observations\label{tab:Observations}}
\tablehead{
			& \colhead{Observation}	& \colhead{$l$}		& \colhead{$b$}		& \colhead{Nominal exposure}	& \colhead{Usable exposure}	& \colhead{$I_{100}$\tablenotemark{a}}	& \colhead{\NH\tablenotemark{b}} \\
\colhead{Observation}	& \colhead{ID}		& \colhead{(deg)}	& \colhead{(deg)}	& \colhead{(ks)}		& \colhead{(ks)}		& (MJy \psr)				& ($10^{20}$ \pcc)
}
\startdata
On filament		& 0084960201		& 278.67		& $-45.32$		& 12.8				& 11.9				& 7.10					& 9.6		\\
Off filament		& 0084960101		& 278.73		& $-47.09$		& 27.8				& 4.4				& 1.22					& 1.9		\\
\enddata
\tablenotetext{a}{100-\micron\ intensity from the all-sky \iras\ maps of \citet{schlegel98}.}
\tablenotetext{b}{Calculated from $I_{100}$ using the conversion relation for the southern Galactic hemisphere in \citet{snowden00}.}
\end{deluxetable*}

Since our current understanding of the particle background of the \xmm\ PN is relatively poor,
and the characterization of the background of the \xmm\ MOS cameras is fairly well refined, we
have restricted our analysis to the data obtained by the two MOS cameras. The data were reduced
as follows. We constructed the light curve in
the 2.5--8.5 \kev\ band for the entire field of view. We fitted a Gaussian to a histogram of the count rates,
and set the ``quiescent level'' to the mean of that Gaussian. We removed from further analysis all
time periods during which the count rate was $>$$3\sigma$ above the quiescent level; the higher
count rate in those time periods is due to either strong soft proton contamination or an enhanced
particle background. Filtering the data using a lower-energy band (0.4--2.0 \kev) produces results no different
from those obtained using the above energy band.
                                                                                
Sources were detected in both the 0.3--2.0 \kev\ and 2.0--10.0 \kev\ bands. Sources with a maximum
likelihood detection value greater than 40 (corresponding to $\sim$$10^{-13}$ \flux) were removed.         
The region removed for each source was a circle whose radius contained 80\%\ of the total flux of a
point source at the source's distance from the optical axis; this radius was typically 24--29 arcseconds. The few
remaining faint point sources are likely to be background AGN with a power-law spectrum which is
unlikely to confuse our analysis of thermal spectra. The actual spectrum of the diffuse emission
was extracted from a region with a radius of 14$\arcmin$ approximately centered on the optical axis
after the removal of the point sources.
                                                                                
We constructed the spectrum of the ``quiescent particle background'' from the ``unexposed corner''
data and filter-wheel-closed data (see \citealp{snowden04}). For each of our two observations, and for each MOS camera,
the background spectrum is modeled using a database of filter-wheel-closed data, scaled by data from
the unexposed corners of the CCDs of that particular camera. This scaling is energy dependent,
and is based upon the hardness and intensity of the unexposed corner data (which varies with time).
The background spectrum is interpolated over the 1.2--1.9 \kev\
interval before being subtracted from the observed spectrum. This region of the spectrum contains
two bright instrumental lines due to aluminum and silicon (at $\sim$1.48 and $\sim$1.74 \kev, respectively).
In most of the spectral fits described below, this region of the spectrum was simply excluded. However,
leaving the instrumental lines in the observed spectrum and fitting them with Gaussians during the analysis
does not significantly affect our results.

The strength of the residual soft proton flares is not known \textit{a priori}, but the shape is reasonably
well modeled by a broken power-law with a break energy of 3.2 \kev, where the spectrum is convolved
with the redistribution matrix but not scaled by the response function.
The contribution of the residual soft proton flares is fitted during the analysis.

\subsection{Solar Wind Charge Exchange}
                                                                                
The solar wind was very steady during both of these observations. The solar proton flux, measured with the
\textit{Advanced Composition Explorer} (\textit{ACE}), was $\sim$$1.8\times10^8$ \pcmsq\ \ps, slightly
below the mean, and the proton speed was 420--440 \kmps, slightly above the mean. The O$^{+7}$/O$^{+6}$ and
O$^{+8}$/O$^{+7}$ ratios had typical values. Both observations were taken at a solar angle of $\sim$$80^\circ$,
avoiding the highest density portions of the magnetosheath. Thus, any solar wind charge exchange (SWCX) contamination
will be relatively low and, more to the point, similar for the two observations.


\section{SPECTRAL MODELING}
\label{sec:Modeling}

\subsection{Spectral Model Description}
\label{subsec:ModelDescription}

The basic model we used to fit to our \xmm\ spectra is based upon that used by \citet{snowden00}
and \citet{kuntz00} in their analyses of RASS data (which is itself a
development of the model used by \citealp{snowden98}). Thus, we use a thermal plasma model in collisional
ionization equilibrium (CIE) for the Local Bubble emission. For the Galactic halo emission we use two thermal plasma components (2$T$),
and for the extragalactic background (due to unresolved AGN) we use a power-law.
In this model, the Local Bubble component is unabsorbed, while the non-local components (halo and extragalactic)
are all subject to absorption. 

We carried out our spectral fitting with XSPEC\footnote{\texttt{http://xspec.gsfc.nasa.gov/docs/xanadu/xspec/}}
v11.3.2p \citep{arnaud96}. Our primary analysis was done using the
Astrophysical Plasma Emission Code (APEC\footnote{\texttt{http://cxc.harvard.edu/atomdb/sources\_apec.html}})
v1.3.1 \citep{smith01a} for the thermal plasma components. For comparison, we also analyzed the spectra using
the \mekal\ model \citep{mewe95}, as discussed in \S\ref{subsec:SpectralModel}. For the absorption we used the \texttt{phabs} model,
which uses cross-sections from \citet{balucinska92}, except for He, in which case the cross-section from \citet{yan98} is used.
Our basic XSPEC model was thus
$\mathtt{apec} + \mathtt{phabs} \ast (\mathtt{apec} + \mathtt{apec} + \mathtt{powerlaw})$.
For chemical abundances we used the interstellar abundance table in \citet{wilms00}\footnote{Implemented using the
XSPEC command \texttt{abund wilm}.}. These abundances were used both by the thermal plasma components and by the
absorption model.

The normalization and the photon index of the power-law used to model the extragalactic background were
frozen at $10.5 (E/\kev)^{-1.46}$ \lineunit\ \pkev\ (\citealp{chen97}; from their model A
fit to \asca\ and \rosat\ data). Note that these values were obtained by \citet{chen97} after removing point
sources down to $\sim$$5 \times 10^{-14}$ \flux, which is roughly equal to the limit to which we have removed
sources (see \S\ref{sec:Observations}). This means that their result should be applicable to our analysis.
Furthermore, the exact values used for the extragalactic power-law have little impact on the fitting
of the thermal emission.

We simultaneously analyzed the on- and off-filament spectra. In the fits the temperatures
and normalizations of all three \texttt{apec} components were free to vary, but were constrained to be the same
for all spectra. The only difference in the model as applied to the different spectra was that the on- and off-filament spectra
had different values of \NH\ for the \texttt{phabs} model. To determine \NH\ for our two pointing directions, we
obtained the 100-\micron\ intensities for these directions from the all-sky \iras\
maps of \citet{schlegel98} and converted them to \NH\ using the conversion relation for the southern Galactic hemisphere
given in \citet{snowden00}. The resulting on- and off-filament column densities are $9.6 \times 10^{20}$ and
$1.9 \times 10^{20}$ \pcmsq, respectively (see Table~\ref{tab:Observations}). Note that the on-filament column
density is consistent with that derived from the color excess of the filament $E(B-V) = 0.17 \pm 0.05$ \citep{penprase98},
which yields $\NH = (8.4 \pm 2.5) \times 10^{20}$ \pcmsq\ when scaled using the conversion relation in \citet{diplas94}.
Also, our results are not very sensitive to the values of \NH\ used.

It should be noted that our absorbing columns do not take into account any contribution from the warm ionized gas
in the Reynolds layer, above the disk of the Galaxy. The average column density of this gas is $7 \times 10^{19} / \sin |b|$~\HII~\pcmsq\ \citep{reynolds91},
which gives 9.6--$9.8 \times 10^{19}$~\HII~\pcmsq\ for our observing directions.
For 1/4-keV X-rays, the effective absorption cross-section per hydrogen nucleus of the ionized gas is 62\%\ of that for neutral
gas \citep{snowden94}, which means that at low energies the Reynolds layer would effectively contribute
an extra $6 \times 10^{19}$~\pcmsq\ to our absorbing columns. However, we find that adding this extra contribution to our absorbing columns affects only the
cooler halo component, and this has no effect on our conclusions. Furthermore, observations of the Galactic absorption towards extragalactic
X-ray sources are generally best fit without the Reynolds layer contribution (e.g.\ \citealp{arabadjis99}).
We therefore proceed using the \iras-derived column densities quoted above.

As stated in \S\ref{sec:Observations}, we also included a broken power-law component in our fit to model the contribution
of residual soft proton flares not removed in the data reduction. This soft proton contamination evidenced itself in certain
\xmm\ spectra as excess emission above that expected from the extragalactic background power-law at energies $\ga 2$ \kev.
The broken power-law parameters were the same for the MOS1 and MOS2 spectra for a given pointing, but were allowed to differ
between the two pointings.

We also experimented with variants of our ``standard'' model. One variation used a non-equilibrium ionization
(NEI) model for the Local Bubble component, i.e.\ we replaced the first \texttt{apec} in our model with the XSPEC \texttt{nei} model.
In this model, the emitting plasma is assumed to have been rapidly heated to some temperature $T$, but the ionization
balance does not yet reflect this new temperature (the ions are underionized). Such a model may be characterized by
an ionization parameter, $\tau = \Ne t$, where \Ne\ is the electron density and $t$ is the time since the heating. 
Collisional ionization equilibrium is reached when $\tau \ga 10^{12}$  \pcc\ \s\ \citep{masai94}.
The \texttt{nei} model is configured to use the Astrophysical Plasma Emission Database (APED)
to calculate the line spectrum\footnote{Implemented using
the XSPEC command \texttt{set neivers 2.0}.}.

Another variation replaced the $2T$ halo model with a model that used a power-law differential emission measure (DEM) of the form
\begin{equation}
	\frac{\mathrm{d}(\mbox{E.M.} [T])}{\mathrm{d}(\log T)} \propto 
	\begin{cases}
		\left( \frac{T}{T_\mathrm{max}} \right)^\alpha	& \text{if $T_\mathrm{min} < T < T_\mathrm{max}$,} \\
		0						& \text{otherwise.}
	\end{cases}
\label{eq:DEM}
\end{equation}
Here the exponent $\alpha$ and the high-temperature cut-off $T_\mathrm{max}$ are free parameters. The low-temperature cut-off
is frozen at $T_\mathrm{min} = 10^5$ \K\ (the results are not very sensitive to the value of $T_\mathrm{min}$, as plasma at this
low a temperature does not significantly contribute to the observed X-ray emission). This model is based upon the XSPEC \texttt{cemekl} model,
though we modified it to use APEC instead of \mekal. The advantage of this model is that it should give a more accurate picture
of the temperature distribution of the gas in the Galactic halo. On the other hand, the $2T$ halo model enables an easier comparison with
the earlier \rosat\ results.

A final variation of our ``standard'' model was to investigate the effect of varying the abundances of various elements.

We simultaneously analyzed the MOS1 and MOS2 spectra obtained from the on- and off-filament pointings.
We used the data between 0.45 and 5 \kev, except for the region between 1.2 and 1.9 \kev, where there are
two bright instrumental lines (see \S\ref{sec:Observations}). An alternative to excluding these data is to
fit these two lines by adding two Gaussians to the model; however, as stated in \S\ref{sec:Observations},
there is no significant difference between the results obtained in these two ways.

In order to better constrain the models at softer energies, we also included two \rosat\ spectra in the fits.
These were extracted from \rosat\ All-Sky Survey data \citep{snowden97} using the HEASARC X-ray Background
Tool\footnote{\texttt{http://heasarc.gsfc.nasa.gov/cgi-bin/Tools/xraybg/xraybg.pl}}
v2.3. The spectra were extracted from 0.5\degr\ radius circles centered on the two \xmm\ pointing
directions, and are normalized to one square arcminute. While larger circles would have reduced the errors on the
\rosat\ data, they would result in the on-filament \rosat\ spectrum being contaminated by off-filament
emission, and vice versa (see Fig.~\ref{fig:FilamentImage}).

We accounted for the difference between the \xmm\ effective field of view and the solid angle used in the \rosat\
extraction by multiplying the model applied to the \xmm\ data by the \xmm\ field of view ($\sim$580 arcmin$^2$),
and normalizing the \rosat\ spectra to 1 arcmin$^2$, as mentioned above.

After our fitting was complete, we converted the XSPEC model normalizations to emission measures
($\int \Ne^2 \dl$) assuming $\nHe / \nH = 0.1$, and neglecting the contribution of metals to the electron density \Ne.

\subsection{A Note on the Abundance Table Used}
\label{subsec:AbundanceTable}

As stated in the previous section, we used the \citet{wilms00} interstellar abundances, 
which differ from those in widely used solar abundance tables \citep[e.g.][]{anders89,grevesse98}
for many astrophysically abundant elements. For example, \citet{wilms00} give an interstellar oxygen
abundance $\nO / \nH = 4.90 \times 10^{-4}$, compared with solar abundances of $8.51 \times 10^{-4}$
\citep{anders89} or $6.76 \times 10^{-4}$ \citep{grevesse98}. However, more recent measurements of the
solar photospheric oxygen abundance \citep{allendeprieto01,asplund04} are in excellent agreement with the
\citet{wilms00} value. Indeed, recent measurements of other metals' photospheric abundances
(e.g.\ C, N, Fe; \citealp{asplund05a}) are lower than the \citet{anders89} and \citet{grevesse98} values,
and are in better agreement with the \citet{wilms00} values. Therefore, in this paper we take the \citet{wilms00}
interstellar abundances to be synonymous with solar abundances.


\section{RESULTS}
\label{sec:Results}

\subsection{Spectral Fit Results}

The results of fitting the above-described ``standard'' ($2T$ halo) model are presented in Table~\ref{tab:FitResults}.
Also shown in this table are the results of using the XSPEC non-equilibrium \texttt{nei} model for the Local Bubble, and of using
a power-law differential emission measure (DEM) for the halo emission (see eq.~[\ref{eq:DEM}]). When we tried varying various chemical
abundances, we found that the deviations from solar abundances were either statistically insignificant or inconsistent
(e.g.\ the Local Bubble oxygen abundance was somewhat dependent upon the halo model used). We therefore do not present
any non-solar-abundance results in Table~\ref{tab:FitResults}, and for the remainder of this paper we just
discuss our solar-abundance results.

\tabletypesize{\scriptsize}
\begin{deluxetable*}{lcccccccccc}
\tablewidth{0pt}
\tablecaption{Spectral fit results\label{tab:FitResults}}
\tablehead{
		&			&					&					&& \multicolumn{5}{c}{$2T$ halo model}																	\\
\cline{6-10}
		& \multicolumn{3}{c}{Local Bubble}									&& \multicolumn{2}{c}{Halo (cool)}					&& \multicolumn{2}{c}{Halo (hot)} 								\\
\cline {2-4} \cline{6-7} \cline{9-10}
		& \colhead{$\log T$}	& \colhead{E.M.\tablenotemark{a}}	& \colhead{$\tau$\tablenotemark{b}}	&& \colhead{$\log T$}		& \colhead{E.M.\tablenotemark{a}}	&& \colhead{$\log T$}	& \colhead{E.M.\tablenotemark{a}} 					\\
\colhead{Model} & \colhead{(K)}		& \colhead{(\emismeas)}			& \colhead{($10^{10}$ \pcc\ \s)}	&& \colhead{(K)}		& \colhead{(\emismeas)}			&& \colhead{(K)}	& \colhead{(\emismeas)}		& \colhead{$\chisq/\mbox{dof}$}
}
\startdata
``Standard'' (CIE Local Bubble)
		& $6.06^{+0.02}_{-0.04}$& 0.018					& \nodata				&& $5.93^{+0.04}_{-0.03}$	& 0.17					&& $6.43 \pm 0.02$	& 0.011				& $435.86/439$				\\
NEI Local Bubble& $6.21^{+0.08}_{-0.10}$& 0.011					& $3.4^{+4.7}_{-1.4}$			&& $5.90^{+0.02}_{-0.03}$	& 0.22					&& $6.44 \pm 0.02$	& 0.010				& $434.56/438$				\\
No Local Bubble	& \nodata		& \nodata				& \nodata				&& $5.92 \pm 0.01$		& 0.37					&& $6.44 \pm 0.02$	& 0.012				& $801.12/441$				\\
One halo component
		& $6.05^{+0.01}_{-0.02}$& 0.023					& \nodata				&& \nodata			& \nodata				&& $6.37 \pm 0.01$	& 0.018				& $539.57/441$				\\
Hot Local Bubble\tablenotemark{c}
		& 6.21 (frozen)		& 0.017					& \nodata				&& $5.29^{+0.13}_{-0.08}$	& 160					&& $6.49 \pm 0.02$	& 0.0073			& $444.04/440$				\\ \\
\hline
\hline
		& \multicolumn{3}{c}{Local Bubble}									&&& \multicolumn{3}{c}{DEM halo model\tablenotemark{d}}																	\\
\cline {2-4} \cline{6-10}
		& \colhead{$\log T$}	& \colhead{E.M.\tablenotemark{a}}	& \colhead{$\tau$\tablenotemark{b}}	&&& \colhead{$\log T_\mathrm{max}$}	&& \colhead{$\alpha$}														\\
\colhead{Model} & \colhead{(K)}		& \colhead{(\emismeas)}			& \colhead{($10^{10}$ \pcc\ \s)}	&&& \colhead{(K)}			&&							&				& \colhead{$\chisq/\mbox{dof}$}		\\
\hline
CIE Local Bubble& $6.02^{+0.01}_{-0.02}$& 0.018					& \nodata				&&& $6.70 \pm 0.07$			&& $-2.01^{+0.14}_{-0.13}$				&				& $438.62/440$				\\
No Local Bubble	& \nodata		& \nodata				& \nodata				&&& $>$6.80				&& $-2.73^{+0.08}_{-0.07}$				&				& $801.64/442$				\\
\enddata
\tablenotetext{a}{Emission measure $\mbox{E.M.} = \int n_e^2 \dl$.}
\tablenotetext{b}{Ionization parameter $\tau = \Ne t$.}
\tablenotetext{c}{See \S\ref{subsec:OtherObs}.}
\tablenotetext{d}{See \S\ref{subsec:ModelDescription} for details of model.}
\end{deluxetable*}

\begin{figure}
\plotone{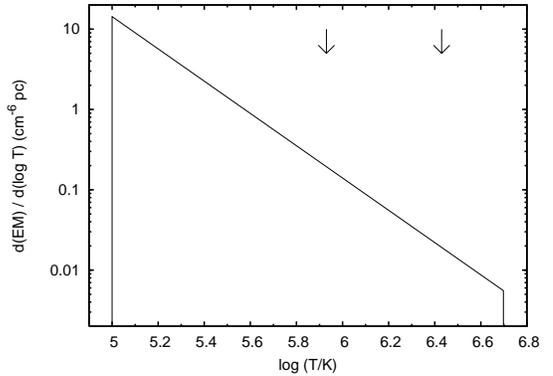}
\caption{Our best-fitting halo power-law differential emission measure (DEM) model. The two arrows mark the temperatures obtained from the $2T$ halo model.\label{fig:DEM}}
\end{figure}

Note from Table~\ref{tab:FitResults} that the Local Bubble model parameters do not significantly change whether one uses
a $2T$ or power-law DEM model for the halo. For completeness we show the halo emission measure given by the
power-law DEM model in Figure~\ref{fig:DEM}, but as this paper is mainly concerned with the Local Bubble, we defer discussion
of the halo results to a later paper. For the remainder of this paper the results discussed will be those obtained
with the $2T$ halo model.

\begin{figure*}
\plottwo{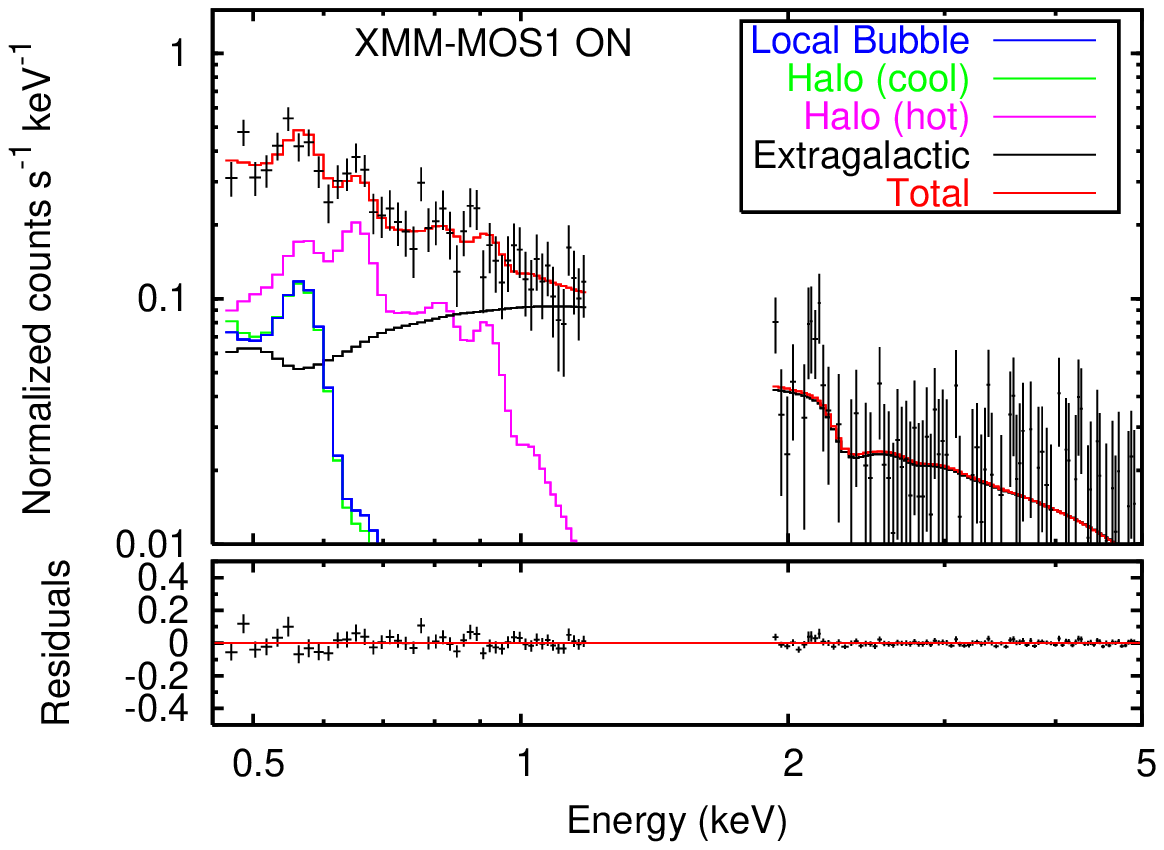}{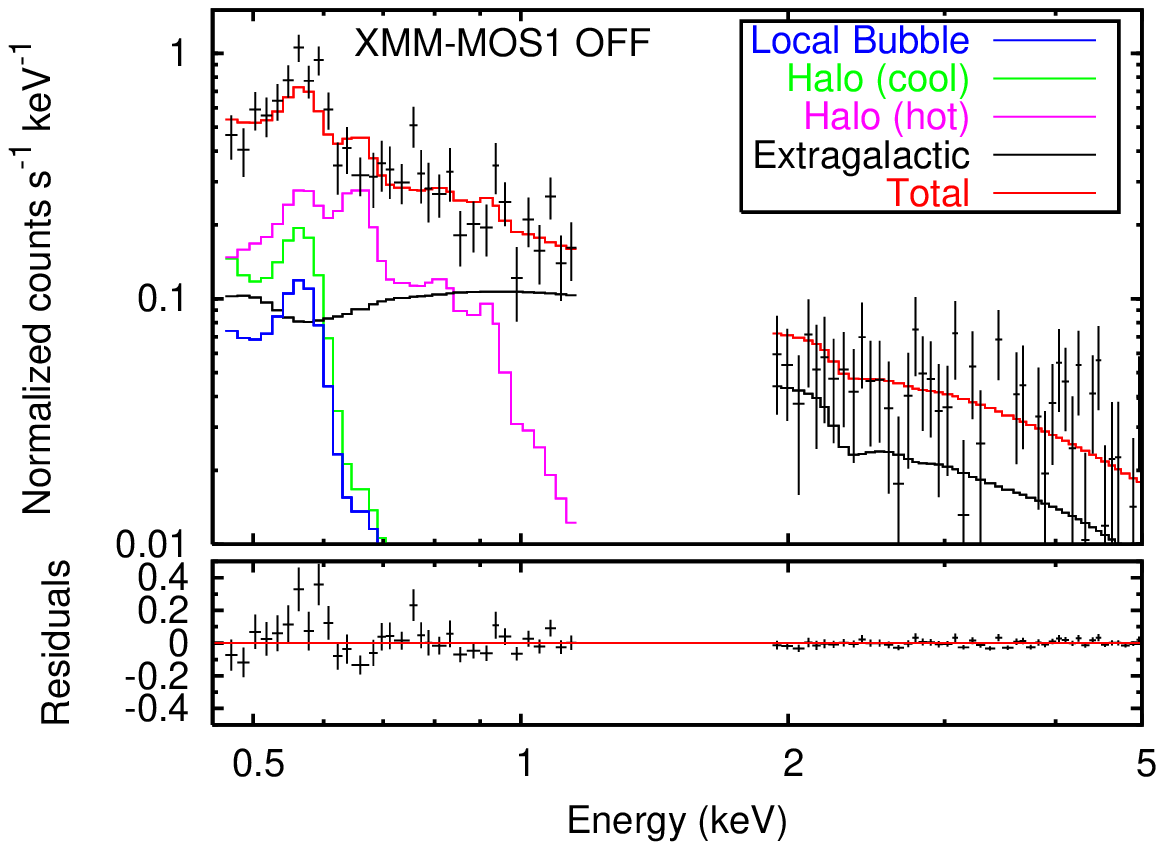} \\
\plottwo{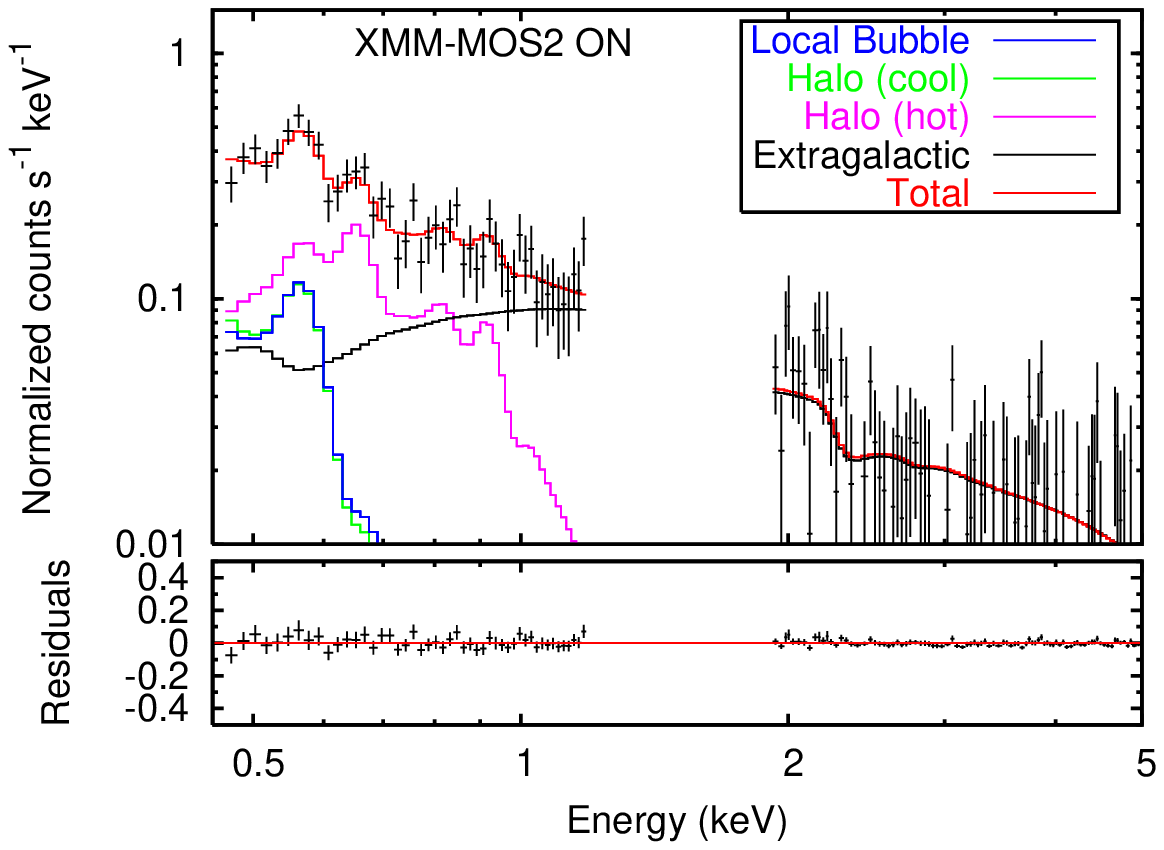}{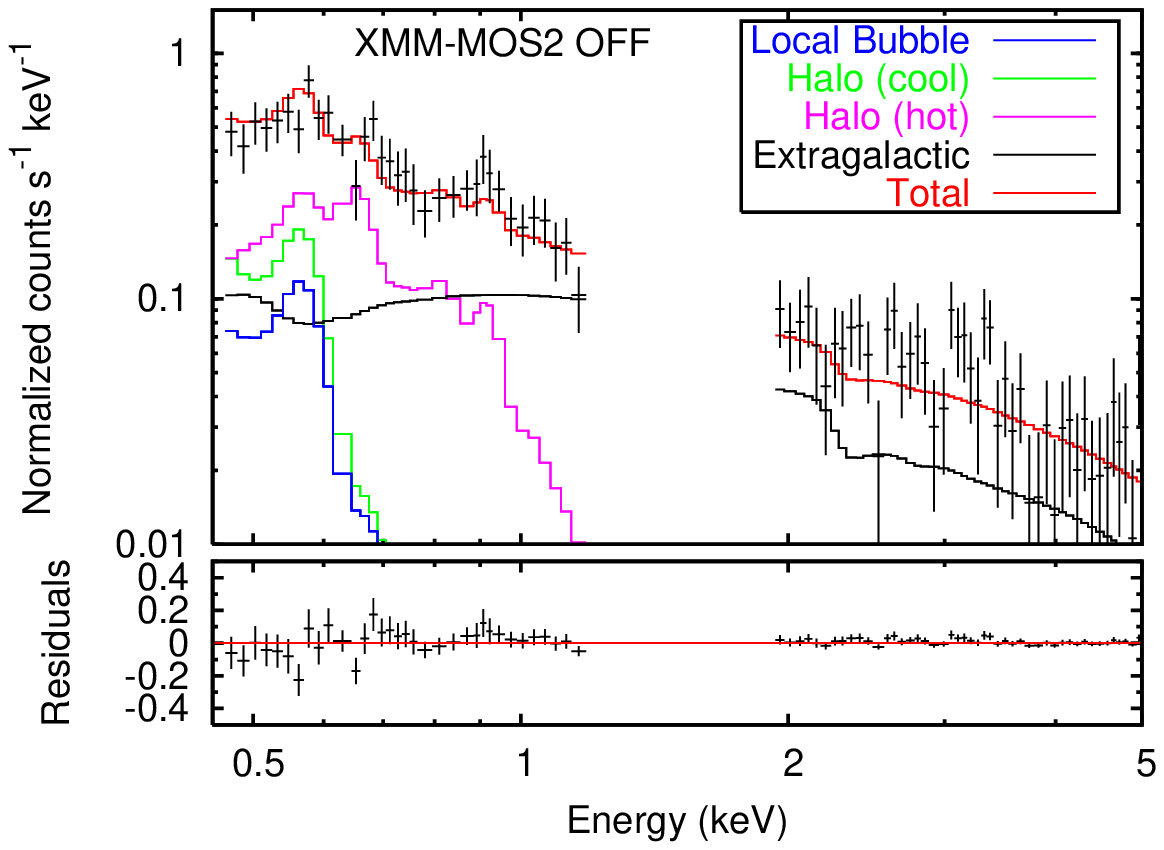} \\
\plottwo{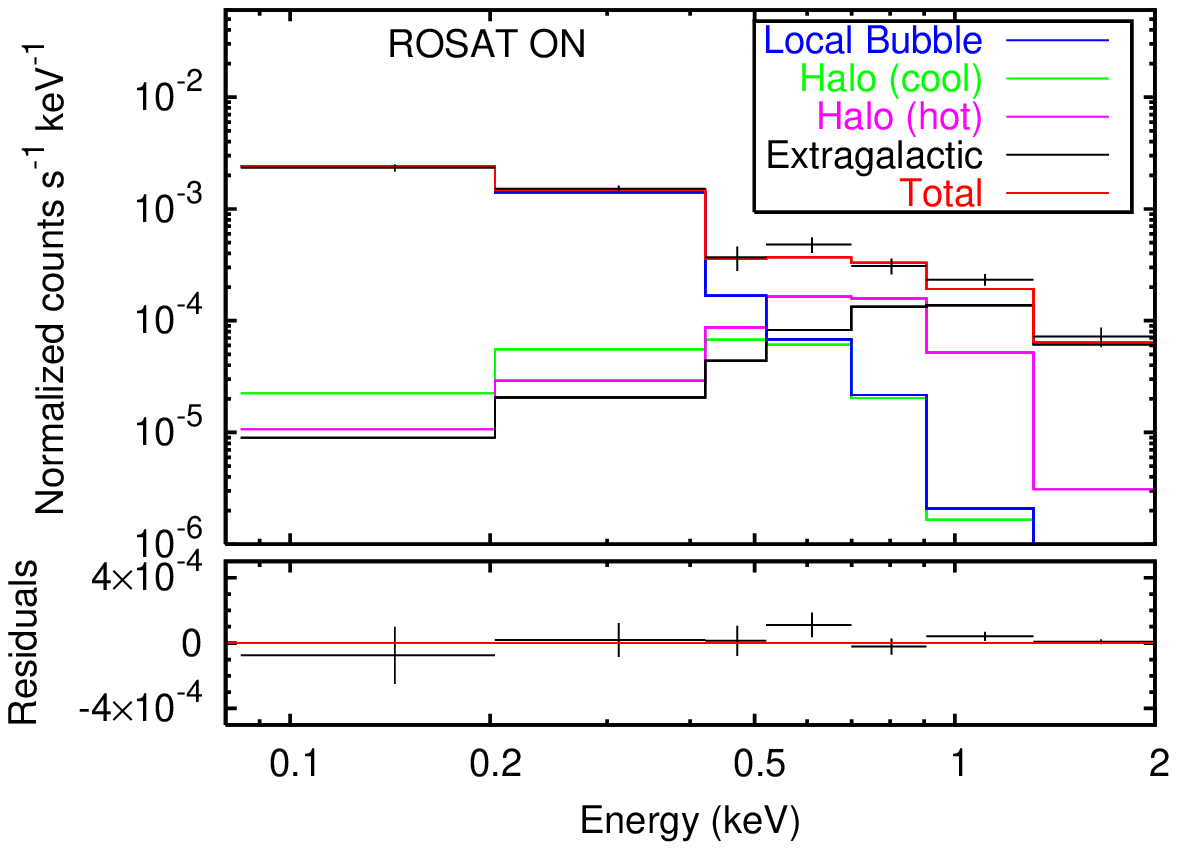}{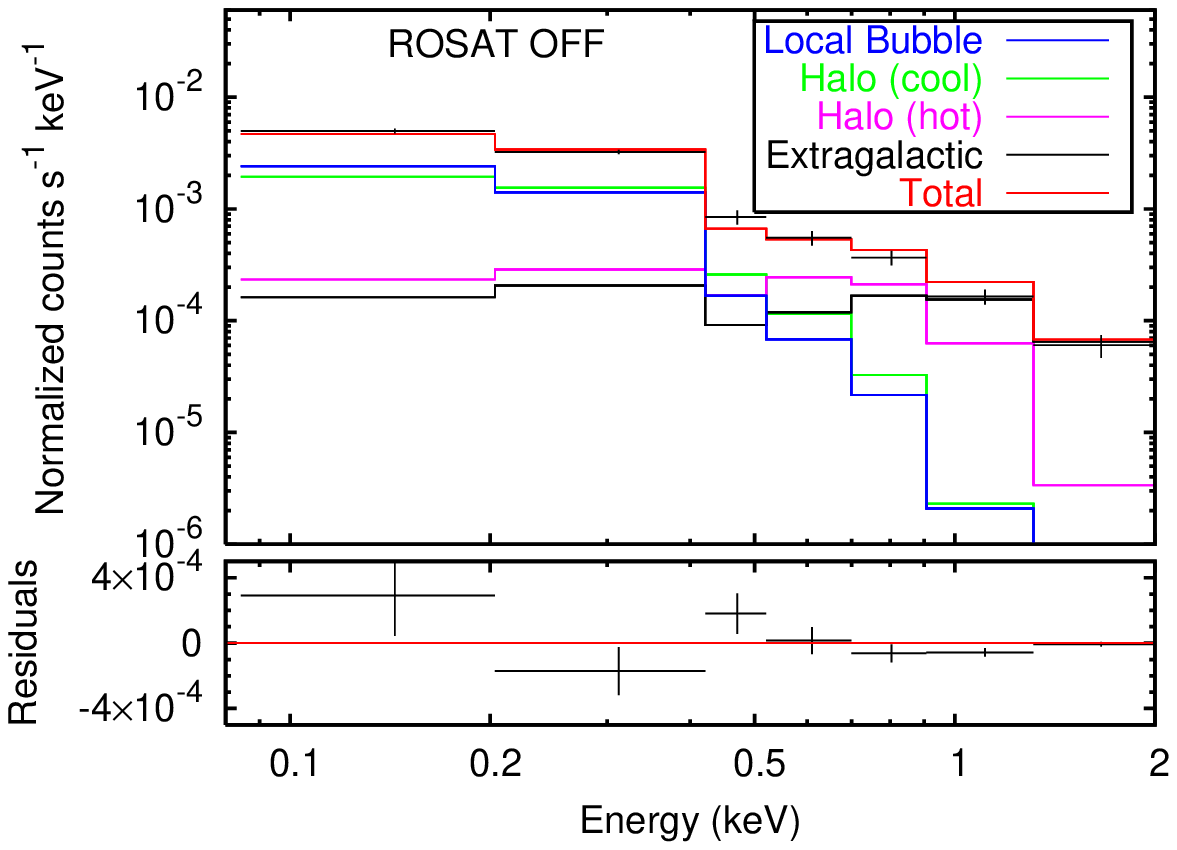}
\caption{Our observed on-filament (left-hand column) and off-filament (right-hand column) spectra, with our best-fit ``standard'' model
(note the different energy ranges on the \xmm\ and \rosat\ plots). The individual components that comprise our model are also illustrated.
Note in the on-filament \xmm\ results that the Local Bubble (blue) and cool halo (green) components overlap. The gap in the \xmm\ data is
where two bright instrumental fluorescence lines have been excluded. The large difference between the \xmm\ and \rosat\ count-rates is because
the \rosat\ data have been normalized to one square arcminute, whereas the \xmm\ data are integrated over the illuminated \xmm\ field
of view ($\approx$580~arcmin$^2$).\label{fig:Spectra}}
\end{figure*}

The spectra and the best-fit ``standard'' model are shown in Figure~\ref{fig:Spectra}. Note the offset between the total model
emission and the extragalactic component above $\sim$2 \kev\ in the off-filament \xmm\ data. This is entirely due to
soft-proton contamination (see \S\ref{subsec:ModelDescription}). Note also that in the \xmm\ band the Local Bubble component makes a significant contribution
only to the \OVII\ emission at $\sim$0.57 \kev. Most ($\sim$90\%) of the \OVIII\ emission at $\sim$0.65 \kev\ is due
to the hotter halo component, with the extragalactic background contributing $\sim$8\%\ of the emission at this
energy. However, the Local Bubble dominates the spectrum at the lowest energies in the on-filament \rosat\ spectrum,
as the halo and extragalactic components are very strongly absorbed. The absorbing cross-section per hydrogen atom
is $8.96 \times 10^{-21}$ \cmsq\ at 0.2 \kev, calculated using the \citet{balucinska92} cross-sections
(except for He; \citealp{yan98}) with the \citet{wilms00} interstellar abundances. This gives an on-filament optical depth of 8.6 at this energy.

While there are some features that are poorly fit in one of the spectra (e.g.\ the model underestimates the \OVII\
emission in the off-filament MOS1 spectrum), when one considers all the spectra together there are no features
that are systematically poorly fit. Overall, the model gives a good fit to the data, with $\rchisq = 0.99$ for 439
degrees of freedom.

We tested whether or not a Local Bubble component is necessary by trying a model without a Local Bubble component. As can be seen from
Table~\ref{tab:FitResults} such a model gives a poor fit to the data whether one uses a $2T$ model ($\rchisq = 1.82$
for 441 degrees of freedom) or a power-law DEM model ($\rchisq = 1.81$ for 442 degrees of freedom) for the halo.
As can also be seen from the table, a $1T$ halo model gives a poor fit to the data ($\rchisq = 1.22$ for 441 degrees
of freedom).

\subsection{Local Bubble O VII and O VIII Intensities}
\label{subsec:OxygenIntensities}

To measure the intensities of the \OVII\ and \OVIII\ emission from the Local Bubble, we replaced the Local Bubble
\texttt{apec} component in our ``standard'' model with a $\mathtt{gaussian} + \mathtt{gaussian} + \mathtt{vapec}$
model, with the oxygen abundance of the \texttt{vapec} component frozen at zero. The two Gaussians model the oxygen line
emission, while the \texttt{vapec} component models the continuum and the contribution of other lines. The widths
of the Gaussians were fixed at zero, and the energies were fixed at 0.5681 and 0.6536 \kev\ for \OVII\ and \OVIII,
respectively. For \OVII\ this is the mean energy of the resonance, intercombination, and resonance lines,
weighted by the line emissivities for a $10^{6.06}$ \K\ plasma. For \OVIII\ it is the weighted mean energy of the
two \Lyalpha\ lines. The relevant line energies and emissivities were obtained from the APEC code
using the XSPEC \texttt{identify} command.
Thawing the energies of the Gaussians did not significantly improve the fit, nor did it
significantly alter the measured intensities. The temperature and normalization of the Local Bubble \texttt{vapec} component
were allowed to vary, while the parameters of all the other components were held fixed at their best-fit values from
the previous section.

We fit this new model to our on- and off-filament \xmm\ and \rosat\ spectra, as before. We measure an \OVII\ intensity
of $3.4^{+0.6}_{-0.4}$ \lineunit, and obtain a $3\sigma$ upper limit on the \OVIII\ intensity of 1.0 \lineunit\
(see Table~\ref{tab:LineIntensities}). We use these results in \S\ref{subsec:Recombining}.

We can check these values using line emissivity data from the ATOMDB database\footnote{\texttt{http://cxc.harvard.edu/atomdb/download.html}\label{footnote:ATOMDB}}.
The temperature and emission measure of our ``standard'' model Local Bubble component yield \OVII\ and \OVIII\ intensities
of 2.9 and 0.017 \lineunit, respectively, including the contribution of dielectronic recombination satellite lines.
These values are consistent with the above-measured values. Although the \OVII\ intensity measured from the Gaussian fits
is higher than that obtained from ATOMDB, this is unlikely to be due to contamination by other oxygen lines, as there
are no other bright oxygen lines in that vicinity. Instead, it is possible that the Local Bubble \texttt{apec} model is slightly
underestimating the contribution of the Local Bubble to the observed \OVII\ emission. This is because this component is not being
constrained solely by the oxygen emission, but also by emission at other energies (e.g.\ the \rosat\ R12 intensity
and the R1/R2 ratio). The \OVIII\ emission is more likely to be contaminated, for example by \OVII\ $n = 3 \rightarrow 1$
emission at 0.6656 \kev. However, this just means that the above upper limit may be overly conservative; it does not adversely
affect our later analysis and conclusions.

\citet{mccammon02} measured the intensities of \OVII\ and \OVIII\ in the soft X-ray background using the X-ray Quantum Calorimeter (XQC),
which was flown on a sounding rocket. They obtained intensities of $4.8 \pm 0.8$ and $1.6 \pm 0.4$~\lineunit\ for \OVII\ and \OVIII,
respectively. These are total intensities for the soft X-ray background averaged over a large area of sky ($\sim$1~sr), whereas
the intensities in Table~\ref{tab:LineIntensities} are just for the Local Bubble. Our results are therefore consistent with the \citet{mccammon02}
results.

\begin{deluxetable}{lcc}
\tablewidth{0pt}
\tablecaption{Local Bubble \OVII\ and \OVIII\ intensities\label{tab:LineIntensities}}
\tablehead{
	& \colhead{Energy}		& \colhead{Intensity}		\\
Ion	& \colhead{(\kev)}		& \colhead{(\lineunit)} 
}
\startdata
\OVII	& 0.5681			& $3.4^{+0.6}_{-0.4}$		\\
\OVIII	& 0.6536			& $<1.0$\tablenotemark{a}	\\
\enddata
\tablenotetext{a}{$3\sigma$ upper limit.}
\end{deluxetable}

\subsection{Derived Parameters of the Local Bubble}

If we assume some spatial extent $L$ for the Local Bubble in our pointing direction, and assume that the Local Bubble
plasma is uniform along the line of sight, we can convert the emission measure found above
to a density. With the measured plasma temperature, this will give us the thermal pressure of the plasma. If we
make the further simplifying assumption that the Local Bubble is a sphere of radius $L$, we can estimate the
thermal energy content and the cooling time of the Local Bubble.

The Local Bubble parameters derived from the ``standard'' model parameters in Table~\ref{tab:FitResults} are
shown in Table~\ref{tab:DerivedParameters}.

\begin{deluxetable}{lc}
\tablewidth{0pt}
\tablecaption{Derived parameters of the Local Bubble\label{tab:DerivedParameters}}
\tablehead{
\colhead{Parameter}		& \colhead{Value}
}
\startdata
Electron density \Ne\ (\pcc)				& $0.013 ( L / 100~\pc )^{-1/2}$		\\
Number density $n$ (\pcc)				& $0.026 ( L / 100~\pc )^{-1/2}$		\\
Pressure $p/k$ (\presalt)				& $2.9 \times 10^4    ( L / 100~\pc )^{-1/2}$	\\
Thermal energy $E_\mathrm{th}$ (\erg)			& $7.4 \times 10^{50} ( L / 100~\pc )^{5/2}$	\\
Cooling time\tablenotemark{a} $t_\mathrm{cool}$ (\yr)	& $1.7 \times 10^7    ( L / 100~\pc )^{1/2}$	\\
Sound crossing time $t_\mathrm{cross}$ (\yr)		& $1.2 \times 10^6    ( L / 100~\pc )$		\\
\enddata
\tablecomments{Calculated from the ``standard'' model parameters for the Local Bubble in Table~\ref{tab:FitResults}, assuming the radius of the Local Bubble is $L$.}
\tablenotetext{a}{Calculated using the Raymond-Smith cooling function (\citealp{raymond77}; \citealp{raymond91}) with \citet{wilms00} abundances:
		$\Lambda (10^{6.06}~\K) = 7.5 \times 10^{-23}$ \erg\ cm$^3$ \ps.}
\end{deluxetable}


\section{DISCUSSION}
\label{sec:Discussion}

In this section, we first compare our results with the results of analyses of \rosat\ All-Sky Survey (RASS) data in \S\ref{subsec:ROSAT}.
In \S\ref{subsec:Halo} we discuss our results in terms of the Galactic halo \OVI\ emission (R.~L. Shelton et~al., in preparation).
We compare our results with those of other \xmm\ and \chandra\ shadowing observations of the Local Bubble in \S\ref{subsec:OtherObs}.
We find a discrepancy between our Local Bubble temperature and that measured by other \xmm\ observations, and discuss
whether or not this can be attributed to our choice of plasma emission code in \S\ref{subsec:SpectralModel}.
Finally we discuss non-equilibrium models of the Local Bubble: in \S\ref{subsec:NEI} we discuss the results of using an underionized (ionizing)
model of the Local Bubble (i.e.\ the XSPEC \texttt{nei} model), and in \S\ref{subsec:Recombining} we discuss our results in terms of
an overionized (recombining) model of the Local Bubble.


\subsection{Comparison with \rosat\ Results}
\label{subsec:ROSAT}

In Table~\ref{tab:ROSAT} we compare our measured temperatures with those measured in various studies of RASS data.
As can be seen, there is excellent agreement between our values and the \rosat-determined values. This is not
surprising, as we use RASS data to constrain our spectral models at low energies. However,
our Local Bubble emission measure ($\int \Ne^2 \dl = 0.018$ \emismeas) is 3--10 times larger than that derived from the \rosat\ data
(0.0018--0.0058 \emismeas; \citealp{snowden98}).

This discrepancy is due to the fact that \citet{snowden98} use the \citet{raymond77} plasma emission code, whereas
we use APEC. Also, \citet{snowden98} assume a higher metallicity in their study: $Z = 0.017$ \citep{allen73} against
$Z = 0.012$ \citep{wilms00}. For a $T = 10^{6.06}$ \K\ plasma, a Raymond \& Smith model with $Z = 0.017$ predicts $\sim$3 times
as much flux in the 0.1--0.5 \kev\ band as an APEC model with $Z = 0.012$. Hence, for a given amount of Local Bubble emission,
our model will give an emission measure $\sim$3 times larger than \citepossessive{snowden98} model, which is what we find.

\begin{deluxetable*}{lccc}
\tablewidth{0pt}
\tablecaption{Comparison with \textit{ROSAT} results\label{tab:ROSAT}}
\tablehead{
					& \colhead{$\log \TLB$}		& \colhead{$\log T_\mathrm{Halo,1}$}	& \colhead{$\log T_\mathrm{Halo,2}$}	\\
\colhead{Work}				& \colhead{(K)}			& \colhead{(K)}				& \colhead{(K)}
}
\startdata
\citet{snowden98}\tablenotemark{a}	& $6.07 \pm 0.05$		& $6.02 \pm 0.08$			& \nodata				\\
\citet{snowden00}			& 6.08\tablenotemark{b}		& 6.00\tablenotemark{b}			& 6.4\tablenotemark{c}			\\
\citet{kuntz00}				& $6.11^{+0.15}_{-0.07}$	& $6.06^{+0.19}_{-0.20}$		& $6.46^{+0.12}_{-0.08}$		\\
This work\tablenotemark{d}		& $6.06^{+0.02}_{-0.04}$	& $5.93^{+0.04}_{-0.03}$		& $6.43 \pm 0.02$			\\
\enddata
\tablenotetext{a}{No second halo component used.}
\tablenotetext{b}{No errors quoted.}
\tablenotetext{c}{Temperature fixed at this value.}
\tablenotetext{d}{``Standard'' model results from Table~\ref{tab:FitResults}.}
\end{deluxetable*}


\subsection{The Galactic Halo O VI Emission}
\label{subsec:Halo}

As already stated, the main focus of this paper is the Local Bubble emission, with discussion of the Galactic halo emission being
deferred to a later paper. However, we note here that our fit results may also be used to make predictions of the halo
\OVI\ intensity, which may then be compared with that measured from an off-filament \textit{Far Ultraviolet Spectroscopic Explorer}
(\fuse) observation. This provides a useful check on our fit results.

Assuming all the halo \OVI\ emission originates from beyond the absorbing material in that direction (which has a
transmissivity to \OVI\ photons of 58\%), the intrinsic intensity of the doublet is
$8070^{+980}_{-1140}$ \lineunit\ (R.~L. Shelton et~al., in preparation). In comparison, the
best-fit temperature and emission measure of the cooler halo component from our ``standard'' model predicts an intrinsic
\OVI\ doublet intensity of $\sim$3000 \lineunit\ (using data from the ATOMDB database\footnote{See footnote \ref{footnote:ATOMDB}.}).
Here we use the $2T$ halo model, as the halo differential emission measure is poorly constrained at low temperatures. Furthermore,
we just consider the cooler halo component, as the contribution of the hotter component to the \OVI\ emission is negligible.

\begin{figure}
\plotone{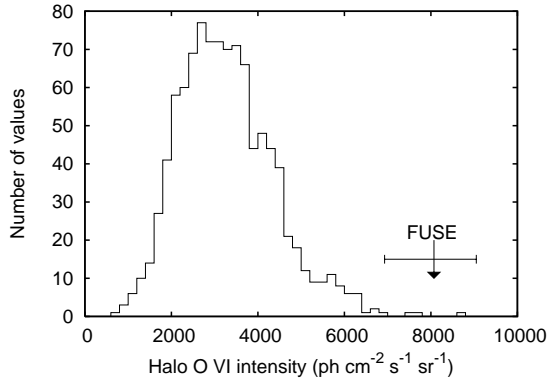}
\caption{Histogram of 1000 randomly generated values of the intrinsic Galactic halo \OVI\ intensity, predicted from our measured values of the
halo temperature and emission measure, taking into account the errors on these parameters. The arrow and the horizontal bar denote the
intrinsic intensity inferred from \fuse, assuming all the \OVI\ emission originates from beyond the absorbing material in that direction
(see text for details).\label{fig:O6IntensityError}}
\end{figure}

We used a Monte Carlo method to estimate the uncertainty on this \OVI\ intensity prediction. We generated 1000 random pairs of
($T$~[halo], E.M.~[halo]), where $\mbox{E.M.~[halo]} = \int \Ne^2 \dl$ is the emission measure of the cooler halo component.
These pairs of numbers were drawn at random from normal distributions whose standard deviations are given by the errors on the
measured parameters in Table~\ref{tab:FitResults} ($\mbox{E.M. [halo]} = 0.17^{+0.05}_{-0.04}$ \emismeas), and were used to
calculate 1000 values of the \OVI\ doublet intensity. We fit a Gaussian to the distribution of these values, and from the mean
and standard deviation obtain a predicted \OVI\ intensity of $3100 \pm 1000$ \lineunit\ (see Fig.~\ref{fig:O6IntensityError}).
The observed intensity is $3.3\sigma$ larger than the predicted intensity (though the distribution of predicted intensities
is positively skewed, which will tend to reduce the significance of this difference). These results indicate that there is more \OVI\
in the halo than is expected from the hot gas alone. This is probably because \OVI\ can also arise in the warm interfaces between cool
clouds and the hot gas.

\subsection{Comparison with Other Shadowing Observations of the Local Bubble}
\label{subsec:OtherObs}

\xmm\ has been used to make shadowing observations of the Local Bubble in the directions of the MBM~12 and Ophiuchus molecular clouds
(\citealp{freyberg03}; \citealp{freyberg04a}), and in the direction of the Bok globule Barnard~68 \citep{freyberg04b}.
\chandra\ has also been used to observe MBM~12 \citep{smith05}.

The Local Bubble temperatures inferred from these \xmm\ observations are consistently higher than the temperature we measure:
$\logTLB = 6.21^{+0.06}_{-0.07}$ for MBM~12 and Ophiuchus (\citealp{freyberg03}; \citealp{freyberg04a}) and
$\logTLB \approx 6.24$ for Barnard~68 (see Appendix), versus $\logTLB = 6.06^{+0.02}_{-0.04}$ from our data.
Note, however, that for MBM~12 there is an additional $\logT \approx 6$ component \citep{freyberg04a}.
It should also be noted that \citet{freyberg03} and \citet{freyberg04a} do not state whether or not they use
\rosat\ data to constrain their fits at lower energies.

In contrast to these \xmm\ results, \citet{smith05} found that they could not explain the \OVII:\OVIII\
ratio in their \chandra\ observation of MBM~12 with an equilibrium Local Bubble model with $\logTLB < 6.3$.
However, they do note the possibility that their \OVIII\ emission is contaminated by emission from another source,
such as solar charge exchange emission.

These other shadowing observations were carried out using much thicker absorbers than our observations:
$4 \times 10^{21}$ \pcmsq\ for MBM~12 \citep{smith05} up to $\sim$$10^{23}$ \pcmsq\ for Barnard~68 \citep{freyberg04b},
against $9.6 \times 10^{20}$ \pcmsq\ for our filament. The optical depth at \OVIII\ energies is at least 2 for these
other observations, implying a transmissivity for background \OVIII\ radiation of less than 14\%. In an observation of one of these
thicker absorbers, one may confidently attribute a large fraction of any observed \OVIII\ emission to the foreground, and thus infer a
higher Local Bubble temperature. However, the higher transmissivity of our filament (61\%\ at \OVIII\ energies) makes it harder to determine
how much of the observed emission is from the Local Bubble, and how much is background emission that has leaked through the filament.
Our best-fit ``standard'' model attributes only $\sim$2\%\ of the observed on-filament \OVIII\ emission to the
Local Bubble, compared with $\sim$30\%\ of the observed \OVII\ emission. The fact that our best-fit model attributes so
little \OVIII\ emission to the Local Bubble leads to our lower value of \TLB. However, the question remains, could more of
the \OVIII\ emission in our observation be due to the Local Bubble? To put this more precisely, are
our data also consistent with $\logTLB = 6.21$?

We tested this by re-fitting our ``standard'' model to the data,  but with
\TLB\ frozen at $10^{6.21}$ \K. The results of this are shown in Table~\ref{tab:FitResults}.
This model does give a good fit to the data ($\rchisq = 1.01$ for 440 degrees of freedom). However, note that the temperature
of the cooler halo component has significantly decreased, and its emission measure has increased by three orders of magnitude. Both
these effects lead to a huge increase in the predicted intrinsic halo \OVI\ intensity (cf.\ \S\ref{subsec:Halo}) to
$\sim$$10^7$ \lineunit. As before, we used a Monte Carlo method to estimate the uncertainty on this prediction. In this case,
there is a much larger error on the emission measure of the cooler halo component ($\mbox{E.M. [halo]} = 160^{+1000}_{-20}$ \emismeas),
resulting in a much larger dynamic range in the predicted intensities. By fitting a Gaussian to the distribution of the logarithms
of the predicted values, we find the predicted intensity is $\log(I_\mathrm{O\,VI} / \lineunitabbr) = 7.5 \pm 1.1$. In comparison,
the intrinsic intensity inferred from the \fuse\ data is $\log(I_\mathrm{O\,VI} / \lineunitabbr) = 3.79^{+0.05}_{-0.06}$
(R.~L. Shelton et~al., in preparation). While we could interpret the discrepancy between the predicted and observed intensities as evidence
that the Galactic halo is out of equilibrium, given the size of the discrepancy (i.e.\ possibly
up to a few orders of magnitude), we instead interpret it as evidence that $\logTLB = 6.21$ is inconsistent with our \xmm\ spectra
and the \fuse\ results when taken together.

It should be noted that two of the clouds discussed above (Ophiuchus and Barnard 68) lie beyond the Local Bubble in the direction of Loop I, near the
Galactic plane. It is therefore possible that the foreground emission (which in the above discussion we have attributed to the Local Bubble) is being
contaminated by emission from Loop I. However, \citet{freyberg04a} states that the weakness of the Fe-L line emission (attributed to Loop I)
in the on-cloud Ophiuchus spectrum indicates that the contamination is small. In contrast to this, MBM~12 does not lie towards any obvious
source of contamination, and may even lie within the Local Bubble, implying that the temperature inferred from these observations is indeed that of
the Local Bubble.

Our pointing direction is $\ga 80^\circ$ away from the directions of the other \xmm\ observations. It is therefore not implausible that our value
of \TLB\ is different from those measured from these observations, as we are observing a different part of the Local Bubble.
These results therefore suggest that the Local Bubble is thermally anisotropic.


\subsection{Choice of Plasma Emission Code}
\label{subsec:SpectralModel}

In this section we consider what effect, if any, the choice of plasma emission code used has on the measured Local Bubble temperature.

The higher foreground temperatures measured from the \xmm\ spectra of MBM~12, the Ophiuchus molecular cloud, and Barnard~68 should
be quite robust, regardless of the plasma emission code used. This is because \OVIII\ emission is observed towards these clouds.
Due to the large optical depths of the clouds, a large fraction of this emission may be attributed to the foreground, implying a higher
temperature than we found towards our filament. \citet{freyberg03} do not give details of the models used in their analysis of the \xmm\
observations of MBM~12 and Ophiuchus, but they do note that $\logTLB = 6.21^{+0.06}_{-0.07}$ ``or higher, depending on the actual model.''
Also, we infer $\logTLB = 6.24$ from \citepossessive{freyberg04b} Barnard~68 data whether we use APEC or \mekal\ (see Appendix). However,
since our filament is less optically thick than these other clouds, there is a greater ambiguity between what is Local Bubble emission and what is
halo emission, and so we test the possibility that different codes would attribute different amounts of the emission in our spectra to the Local Bubble and
the halo, leading to a different value of \TLB.

\begin{figure}
\plotone{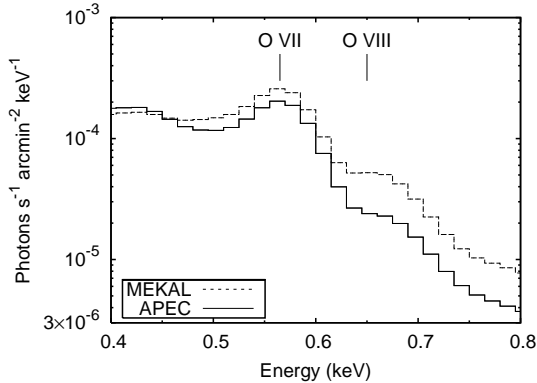}
\caption{Our best fitting APEC and \mekal\ Local Bubble models, folded through the \xmm\ MOS1 instrumental response. Note that there is more Local Bubble \OVIII\
emission at $\sim$0.65 \kev\ in the \mekal\ model than in the APEC model.\label{fig:APECversusMEKAL}}
\end{figure}

We tested this by repeating our fits using the \mekal\ code instead of APEC for the thermal emission components. The results are compared with our
APEC results in Table~\ref{tab:APECandMEKAL}. Note in particular that \mekal\ gives a higher Local Bubble temperature than APEC:
$\logTLB = 6.17^{+0.06}_{-0.07}$ (\mekal) versus $6.06^{+0.02}_{-0.04}$ (APEC).
It should be emphasized that the difference between the APEC and \mekal\ results for our data
is not because the two codes give different temperatures for the same Local Bubble spectrum.
Instead it is because \mekal\ attributes more of the \OVIII\ emission to the Local Bubble (see Fig.~\ref{fig:APECversusMEKAL}), and
correspondingly less to the halo. Hence, the inferred Local Bubble spectrum is different between the two codes, and thus so too is \TLB.
This discrepancy is most likely to be due to uncertain modeling of the lines from L-shell ions of Ne, Mg, and Si, which dominate
the emission at the lowest \rosat\ energies. The ATOMDB v1.3.1 release
notes\footnote{\texttt{http://cxc.harvard.edu/atomdb/issues\_improvements.html}} contain a caveat that there are very few data on lines from
these ions (other than Li-like ions). Differences between the codes in this energy regime would affect the fits to the \rosat\
data, which would then affect the fitting to the higher-energy \xmm\ data.

To test whether or not this does cause the discrepancy between the APEC and \mekal\ results, we re-fit our ``standard'' model just to the \xmm\
spectra, without the \rosat\ data. These results are also shown in Table~\ref{tab:APECandMEKAL}. Note that these temperatures are unphysical,
as they are inconsistent with the low-energy \rosat\ data. However, the agreement between the codes' results is much better than before,
suggesting that at least part of the discrepancy is due to uncertain modeling of lines in the lowest-energy \rosat\ bins.

Despite this discrepancy, we reiterate that the measurement of $\logTLB \ge 6.21$ from the other \xmm\ observations should not be strongly code-dependent,
and note that we obtain a lower temperature than this whether we use APEC or \mekal. This therefore implies that the conclusion of the previous section, namely that the
Local Bubble appears to be thermally anisotropic, is not an artefact of our choice of plasma emission code. However, we should reiterate the possibility that
the foreground emission in some of the other \xmm\ observations may be contaminated by non-Local Bubble emission.

\begin{deluxetable*}{lccccc}
\tablewidth{0pt}
\tablecaption{Comparison of APEC and \mekal\ results\label{tab:APECandMEKAL}}
\tablehead{
		& \colhead{\rosat}		& \colhead{$\log \TLB$}		& \colhead{$\log T_\mathrm{Halo,1}$}	& \colhead{$\log T_\mathrm{Halo,2}$}	\\
\colhead{Code}	& \colhead{included?}		& \colhead{(K)}			& \colhead{(K)}				& \colhead{(K)}				& \colhead{$\chisq/\mbox{dof}$}
}
\startdata
APEC		& Y				& $6.06^{+0.02}_{-0.04}$	& $5.93^{+0.04}_{-0.03}$		& $6.43 \pm 0.02$			& $435.86/439$	\\
\mekal		& Y				& $6.17^{+0.06}_{-0.07}$	& $5.67^{+0.01}_{-0.02}$		& $6.35 \pm 0.02$			& $433.68/439$	\\
APEC		& N 				& $6.44^{+0.03}_{-0.08}$	& $6.20^{+0.04}_{-0.11}$		& $6.69^{+0.10}_{-0.13}$		& $410.21/425$	\\
\mekal		& N 				& $6.37^{+0.06}_{-0.10}$	& $6.21^{+0.13}_{-0.07}$		& $6.75^{+0.15}_{-0.20}$		& $411.57/425$	\\
\enddata
\end{deluxetable*}


\subsection{Ionizing Model of the Local Bubble}
\label{subsec:NEI}

In the preceding sections we have just discussed collisional ionization equilibrium (CIE) plasma
models. In this section we discuss the results of using the XSPEC \texttt{nei} model to model the Local Bubble emission
(see Table~\ref{tab:FitResults}).
This is a model in which the ions are underionized, i.e.\ the plasma has been rapidly
heated, but the ionization balance does not yet reflect the new temperature: the ions are still in the process
of ionizing. As stated in \S\ref{subsec:ModelDescription}, such a model may be characterized by an ionization
parameter, $\tau = \Ne t$, where \Ne\ is the electron density and $t$ is the time since the heating. 
Equilibrium is reached when $\tau \ga 10^{12}$ \pcc\ \s\ \citep{masai94}.

In certain situations one may use the \ftest\ to determine whether or not adding an extra model parameter leads to a
significant improvement in \chisq\ \citep{bevington03}. Unfortunately we cannot use the \ftest\ here to test whether
or not the non-equilibrium model is an improvement on the equilibrium model. This is because the additional parameter ($\tau$)
is on the boundary of the set of possible parameter values in the simpler (null) model (i.e.\ $\tau = \infty$ in the
equilibrium model). In such a case, one cannot use the \ftest\ \citep{protassov02}.

However, we can note that the difference in \chisq\ between the two models is very small, suggesting that thawing $\tau$
does not lead to a significant improvement in \chisq\ (even if we cannot use the \ftest\ to demonstrate this for certain).
On the other hand, it should be noted that the value of $\tau$ we measure ($[3.4^{+4.7}_{-1.4}] \times 10^{10}$ \pcc\ \s)
is significantly less than what is expected in the equilibrium case ($\tau \ga 10^{12}$ \pcc\ \s; \citealp{masai94}), which
means that our data cannot rule out an ionizing non-equilibrium model.

In order to distinguish between equilibrium and non-equilibrium models we would require more spectral information.
With our data, the only Local Bubble ``line'' (actually a blend of lines) we can clearly see is
the \OVII\ feature at $\sim$0.57 \kev. There are no other obvious Local Bubble lines in the \xmm\ bandpass
(the observed \OVIII\ feature in the on-filament spectrum is attributed to leak through from the Galactic halo),
and we only have two \rosat\ spectral bins at energies below the \xmm\ bandpass. In contrast to this, one
needs to compare the flux ratios of lines from several different ions in order to determine whether or not
non-equilibrium effects are important.

In principle, we could use data from other wavebands to help distinguish between the two classes of model.
\fuse\ has observed the Local Bubble in the same directions
as our \xmm\ observations. Using the on-filament observation, \citet{shelton03} has placed
a $2\sigma$ upper limit on the Local Bubble \OVI\ $\lambda\lambda1032$, 1038 doublet intensity of 800 \lineunit\
(including statistical and systematic uncertainties).
In comparison, the predicted \OVI\ intensities of our best-fit models (determined using XSPEC) are
180 \lineunit\ (equilibrium) and 63 \lineunit\ (non-equilibrium), both of which are consistent with the observed upper limits.

We therefore conclude that our data are unable to distinguish between an equilibrium plasma model and a non-equilibrium
model in which the ions are underionized and are in the process of ionizing.


\subsection{Recombining Model of the Local Bubble}
\label{subsec:Recombining}

In this section we consider an alternative non-equilibrium Local Bubble model to the previous section, namely one in which the plasma
is overionized and the ions are in the process of recombining.

Such a model for the Local Bubble was originally proposed by
\citeauthor{breitschwerdt94} (\citeyear{breitschwerdt94}; see also \citealp{breitschwerdt96a}; \citealp{breitschwerdt96b}; \citealp{breitschwerdt01}).
In this model, the Local Bubble is assumed to be the relic of an old superbubble formed by $\sim$10 successive supernovae
in a dense ($\sim$$10^4$ \pcc) molecular cloud. A few million years ago, the bubble burst out of the cloud and underwent
rapid adiabatic expansion into the surrounding medium, assumed to be $\sim$10--100 times less dense than the cloud.
During this process the bubble adiabatically cooled to $\la 10^5$ \K\ (i.e.\ much less than the temperature inferred from
the X-rays assuming collisional ionization equilibrium). However, owing to the rapidity of the cooling, the ions remained in
high ionization states. The observed X-rays are then supposed to be due to the delayed recombinations of these overionized
ions, rather than being from a hot ($\sim$$10^6$ \K) plasma in equilibrium.

A recombining plasma model has a number of appealing features. Firstly, there is an order of magnitude difference between the pressure of
the Local Cloud in which the solar system resides ($P_\mathrm{LC} / k \sim 2000$ \presalt; \citealp{lallement98}) and the
pressure of the surrounding Local Bubble inferred from the X-ray data assuming CIE
($P_\mathrm{LB} / k \sim$ 15,000 \presalt\ \citepsq{snowden98} to 29,000 \presalt\ [Table~\ref{tab:DerivedParameters}]). It is difficult to
see how such a large pressure difference can be maintained. The lower Local Bubble temperature in the \citet{breitschwerdt94} model greatly reduces the Local Bubble
pressure, and thus eliminates this problem. Secondly, the electron density inferred from the dispersion measure of the pulsar
PSR~$0950 + 08$ ($d \approx 130$ \pc) is 0.023 \pcc\ \citep{reynolds90}. If this density is representative of the Local Bubble,
then a $\sim$$10^6$ \K\ plasma in equilibrium would produce too much X-radiation (cf.\ the electron density inferred from the X-ray data
assuming CIE is $\sim$0.007 \pcc\ \citepsq{snowden98} to 0.013 \pcc\ [Table~\ref{tab:DerivedParameters}]). However, a
higher \Ne\ is acceptable within the \citet{breitschwerdt94} model. Thirdly, a $\sim$$10^6$ \K\ plasma in equilibrium should produce
most of its emission as lines in the extreme ultraviolet (EUV). However, such emission is not observed.
\textit{Extreme Ultraviolet Explorer} (\textit{EUVE}) spectra of the diffuse EUV background ($\sim$17--80 \ev)
place an upper limit on the emission measure of a local $10^6$ K plasma which is an order of magnitude less than the Local Bubble emission
measures in Table~\ref{tab:FitResults} \citep{jelinksy95,vallerga98}.
A recombining plasma model can explain this observation, as the much lower kinetic temperature suppresses collisional excitation
of the EUV lines \citep{breitschwerdt96a}.

Here we examine whether or not a recombining plasma model can simultaneously explain our observed Local Bubble \OVII\ intensity and
the observed upper limits on the \OVI\ and \OVIII\ intensities. We do this by first calculating the fractions of oxygen in
the ionization states \OVI\ to \OIX, and then using these populations to predict the line intensities. These are then
compared with the observed limits. For this purpose we use $3 \sigma$ limits.  Note that \citet{shelton03} does not explicitly quote a
$3\sigma$ upper limit for the \OVI\ intensity, but we can obtain one from the data in her Table~2. The $3\sigma$ upper limit on
the $\lambda1032$ intensity obtained from the combined $\mbox{day} + \mbox{night}$ \fuse\ data is 710 \lineunit\ (statistical
uncertainty only). As the theoretical $\lambda1032$-to-$\lambda1038$ intensity ratio is 2 to 1, the $3\sigma$ upper limit on the
doublet intensity is 1065 \lineunit.

In a purely recombining plasma (i.e.\ one in which ionizations can be ignored), the fraction $p_i$ of the atoms of a given
element in an ionization state with $i$ electrons is given by
\begin{equation}
	\frac{\dd p_i}{\dd t} = -R_i \Ne p_i + R_{i-1} \Ne p_{i-1}
\label{eq:RateEquation}
\end{equation}
where $R_i$ is the recombination coefficient (radiative + dielectronic)
to go from the $i$th state to the ($i+1$)th state (e.g.\ $R_0$ is the recombination
coefficient for \OIX\ $\rightarrow$ \OVIII). Note that for $i = 0$ there is no second term in equation~(\ref{eq:RateEquation}),
and it may easily be solved for $p_0$. This solution may then be used to solve equation~(\ref{eq:RateEquation}) for $p_1$, and so on.
\citet{smith05} give the solutions for this system of coupled differential equations for the case of a plasma that is initially fully
ionized, i.e.\ $p_0 = 1$, $p_{i > 0} = 0$ at $t = 0$ (their eqs.~[5] and [6]). We used these solutions
to calculate the populations of \OIX\ to \OVI, using radiative recombination
coefficients from \citet{verner96} (the exception to this is \OVI\ $\rightarrow$ \OV, for which \citeauthor{verner96} do
not give data; in this case we use data from \citealp{shull82}), and dielectronic recombination coefficients from \citet{mazzotta98}.

To calculate the \OVII\ and \OVIII\ intensities, we use rate coefficients $\alpha^\mathrm{R}$ for the formation of the lines of interest
due to recombinations (radiative + dielectronic) from the above ionization stage, calculated using data from \citet{mewe85}. We can 
neglect the contribution of collisional excitations to the lines, as the temperature of the cooled plasma ($kT \la 20$ \ev) is much
less than the energy required to excite an electron to the $n = 2$ level ($\sim$570 \ev). The intensity \Iovii\ of \OVII\ emission
is therefore given by
\begin{eqnarray}
	\Iovii	&=& \frac{\alpha^\mathrm{R}_1}{4\pi} \int \Ne \noviii \dl				\nonumber \\
		&=& \left( \frac{\alpha^\mathrm{R}_1 p_1}{4.8\pi} \right) \left( \frac{\nO}{\nH} \right) \int \Ne^2 \dl,
\label{eq:IOVII}
\end{eqnarray}
where $\alpha^\mathrm{R}_1$ is the rate coefficient for the production of \OVII\ \Kalpha\ (i.e.\ $n = 2 \rightarrow 1$)
due to recombinations from \OVIII, comprising contributions from the resonance, intercombination, and forbidden lines
(the contribution of unresolved satellite lines is negligible in this model; \citealp{mewe81}).
In equation~(\ref{eq:IOVII}) we have used
\begin{eqnarray}
	\noviii &=& \left( \frac{\noviii}{\nO} \right) \left( \frac{\nO}{\nH} \right) \left( \frac{\nH}{\Ne} \right) \Ne \nonumber \\
		&=& p_1 \left( \frac{\nO}{\nH} \right) \left( \frac{\Ne}{1.2} \right),
\label{eq:OIXfraction}
\end{eqnarray}
where $\noviii/\nO \equiv p_1$ and $\nH/\Ne = 1/1.2$ for a fully ionized plasma with $\nH / \nHe = 10$. Similarly, the intensity \Ioviii\ of \OVIII\
emission is given by
\begin{equation}
	\Ioviii = \left( \frac{\alpha^\mathrm{R}_0 p_0}{4.8\pi} \right) \left( \frac{\nO}{\nH} \right) \int \Ne^2 \dl,
\end{equation}
where $\alpha^\mathrm{R}_0$ is the rate coefficient for the production of \OVIII\ \Lyalpha\ due to recombinations from \OIX.

For \OVI, collisional excitation dominates over recombinations from \OVII. The intensity of the $\lambda\lambda1032$, 1038 doublet
may be calculated by rearranging equation~(5) from \citet{shull94}:
\begin{eqnarray}
	\Iovi	&=&	\frac{\bar{\Omega}(T)}{\sqrt{T_5}} \exp \left( - \frac{1.392}{T_5} \right) \nonumber \\
		&\times&\left( \frac{\int \Ne \novi \dl}{9.21 \times 10^8 ~\cm^{-5}} \right) ~\lineunitabbr,
\end{eqnarray}
where $T_5$ is the gas temperature $T$ in units of $10^5$ \K, and $\bar{\Omega} (T)$ is the Maxwellian-averaged collision strength
for de-excitation of the \OVI\ doublet. We use the parameterization for $\bar{\Omega} (T)$ given by equation~(6) in
\citet{shull94}. We calculate \novi\ from \Ne\ using the oxygen abundance and $p_3 \equiv \novi / \nO$:
$\novi = p_3 (\nO / \nH) (\Ne/1.2)$ (cf.\ eq.~[\ref{eq:OIXfraction}]).

\begin{figure}
\plotone{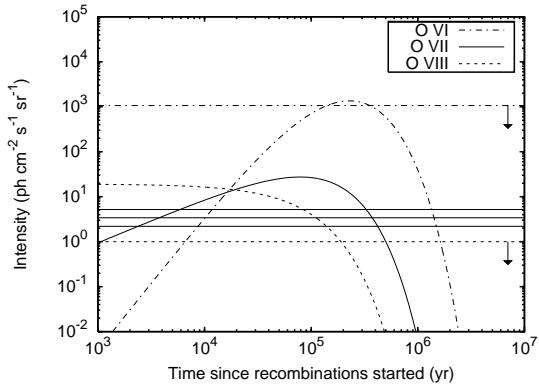}
\caption{\OVI, \OVII, and \OVIII\ intensities predicted by our recombining plasma model as a function of time
(the plasma is fully ionized at $t = 0$). The curves were calculated using $\Ne = 0.024$ \pcc\ and $T = 4.3 \times 10^4$ \K\
\citep{breitschwerdt01}. The horizontal lines show our observed Local Bubble \OVII\ intensity
(with the $3\sigma$ uncertainty; see Table~\ref{tab:LineIntensities}), and the $3\sigma$ upper limits on the intensities of \OVI\
\citep{shelton03} and \OVIII\ (see Table~\ref{tab:LineIntensities}).
\label{fig:RecombiningModel}}
\end{figure}

We have calculated oxygen line intensities as functions of the time $t$ since recombinations began
for electron densities between 0.001 and 0.3 \pcc\ and temperatures
between $10^4$ and $10^7$ \K. Figure~\ref{fig:RecombiningModel} shows an example of our results, calculated for $\Ne = 0.024$ \pcc\
and $T = 4.3 \times 10^4$ \K\ \citep{breitschwerdt01}. For a given (\Ne,~$T$), the tightest constraint on $t$
is set by the measured \OVII\ intensity. The predicted \OVI\ and \OVIII\ intensities at this time may then be compared with the
observed upper limits. For example, in Figure~\ref{fig:RecombiningModel} the observed \OVII\ intensity implies that
$t \sim \mbox{few} \times 10^3$ or $\sim$3--$4 \times 10^5$ \yr. The former case is strongly ruled out as the predicted
\OVIII\ intensity is an order of magnitude larger than the observed $3\sigma$ upper limit. In the latter case, however, both the \OVI\ and the
\OVIII\ intensities are within the observed $3\sigma$ upper limits.

\begin{figure}
\plotone{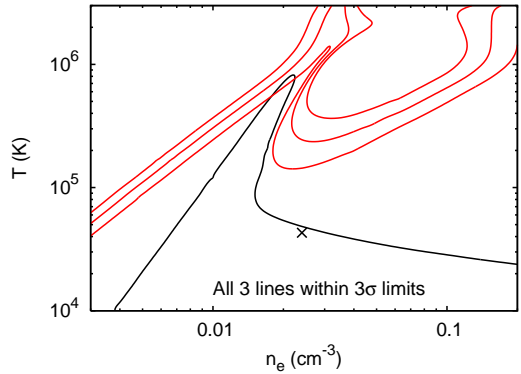}
\caption{\textit{Black contour}: for values of (\Ne,~$T$) below this contour there exists a time for which all
three predicted line intensities simultaneously lie within the observed $3 \sigma$ limits. Above this contour no such time exists.
\textit{Red contours}: contours showing the time $t$ after recombinations began at which the predicted \OVII\ intensity is
closest to the observed value (from bottom to top, $t = 6$, 7, and $8 \times 10^5$ \yr).
\textit{Black cross}: This marks $\Ne = 0.024$ \pcc, $T = 4.3 \times 10^4$ \K\ (\citealp{breitschwerdt01}; see Fig.~\ref{fig:RecombiningModel}).
\label{fig:RecombiningModelChisq}}
\end{figure}

For each (\Ne,~$T$) we investigated, we check whether or not there exists a time (more precisely, a range of times) for which all
three predicted line intensities lie within their observed $3 \sigma$ limits. The results are illustrated by the black contour in 
Figure~\ref{fig:RecombiningModelChisq} -- in the region above this contour, no time exists for which all three lines are
simultaneously within their observed limits. In the left portion of this region, the \OVII\ intensity nevers reaches the observed
value, while in the right portion, either the \OVI\ or the \OVIII\ is too bright whenever the \OVII\ is within the observed limits.
We note that this is not a rigorous statistical test. Such a test would be difficult, if not impossible, as we only have three
data points, two of which are upper limits (so, for example, the \chisq\ test cannot be used). However, it does help to illustrate
which regions of (\Ne,~$T$)-space are and are not observationally acceptable.

Also for each (\Ne,~$T$) we find the time for which the predicted \OVII\ intensity is equal to the observed value (or closest to it, if it
never actually equals it). This provides the tightest constraint on the time $t$ since recombinations began. If there are two such values
(as in Figure~\ref{fig:RecombiningModel}) we take the later one. These results are illustrated by the red contours in 
Figure~\ref{fig:RecombiningModelChisq}. As can be seen, the observationally acceptable region of (\Ne,~$T$)-space (i.e.\ below the black
contour) corresponds to $t < 7 \times 10^5$ \yr, and most of this region (including the values of \Ne\ and $T$ from \citealt{breitschwerdt01})
corresponds to $t < 6 \times 10^5$ \yr. Furthermore, the observationally acceptable region above the $6 \times 10^5$-\yr\ contour corresponds to
$T \ga 5.5 \times 10^5$ \K\ and $\Ne \ga 0.018$ \pcc. This gives a pressure $P \ga 20000$ \presalt, which is an order of magnitude larger than
the pressure of the Local Cloud \citep{lallement98}. While this is not as strong a constraint as the observed oxygen line intensities, it does
mean that in this region of parameter space the recombining model loses one of its main selling points.

In summary, for most of the observationally acceptable region of (\Ne,~$T$)-space, the observed \OVII\ intensity implies that the time since
recombinations began is $t < 6 \times 10^5$ \yr. We shall now show that this time is also an upper limit on the time since the Local Bubble burst out
of its natal cloud, as the timescale for recombinations was shorter when the Local Bubble was smaller.

If the Local Bubble is expanding and cooling adiabatically, its temperature $T$ and volume $V$ are related by
\begin{equation}
	TV^{\gamma-1} = \mathrm{constant},
\end{equation}
where $\gamma$ is the adiabatic index ($\gamma = 5/3$). Therefore, the temperature of the Local Bubble when it is of linear size
$L$ is
\begin{equation}
	T = T_0 (L / L_0)^{-2},
\end{equation}
where $T_0$ and $L_0$ are the present-day temperature and size. Similarly, the electron density \Ne\ is related to
the present-day density ${\Ne}_0$ by
\begin{equation}
	\Ne = {\Ne}_0 (L / L_0)^{-3}.
\end{equation}
The recombination timescale $t_\mathrm{rec}$ is defined by
\begin{equation}
	t_\mathrm{rec} \equiv \frac{1}{\Ne R(T)},
\end{equation}
where $R(T)$ is the recombination coefficient (see eq.~[\ref{eq:RateEquation}]). This timescale may be expressed as a function of
$L/L_0$,
\begin{equation}
	\frac{t_\mathrm{rec}}{t_\mathrm{rec,0}} = \frac{ (L/L_0)^3 R(T_0) }{ R( T_0 [ L/L_0 ]^{-2} ) },
\end{equation}
where $t_\mathrm{rec,0}$ is the present-day recombination timescale.

\begin{figure}
\centering
\includegraphics[width=0.75\linewidth]{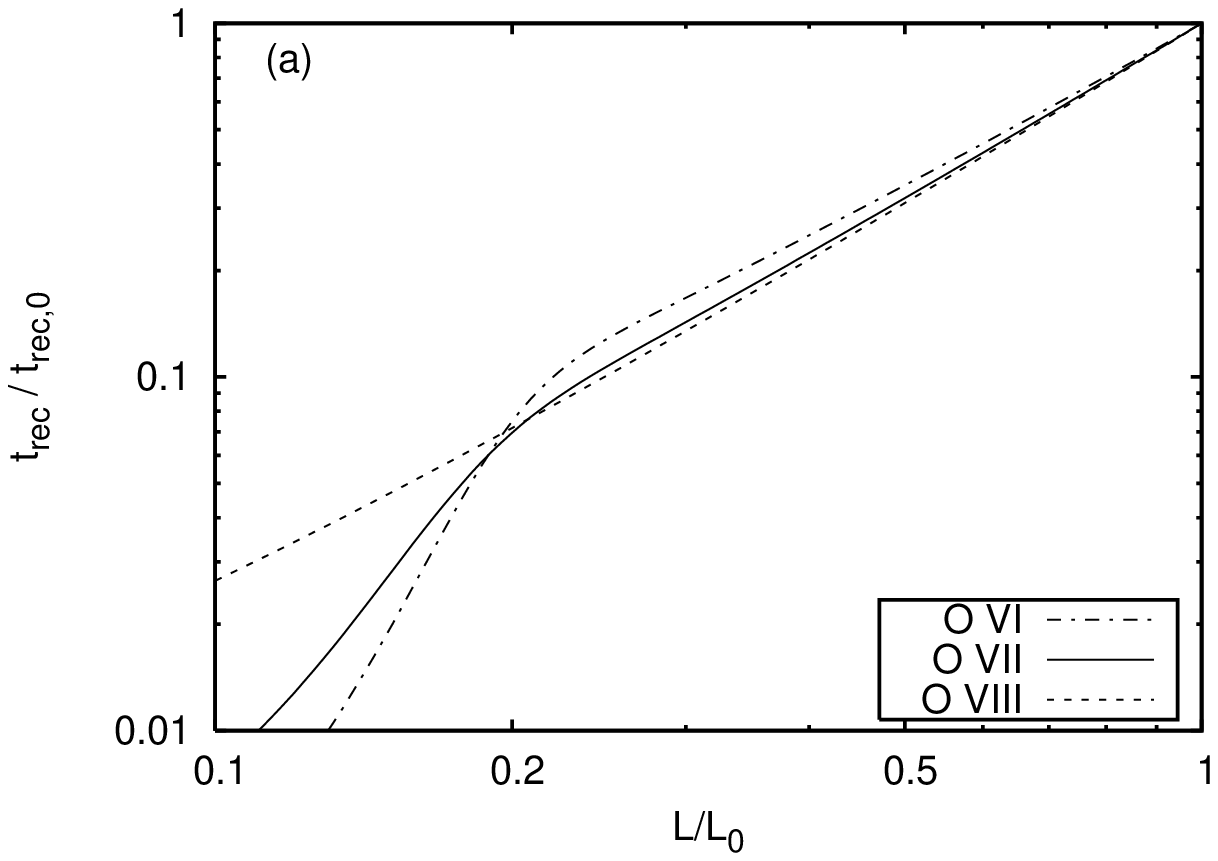} \\
\includegraphics[width=0.75\linewidth]{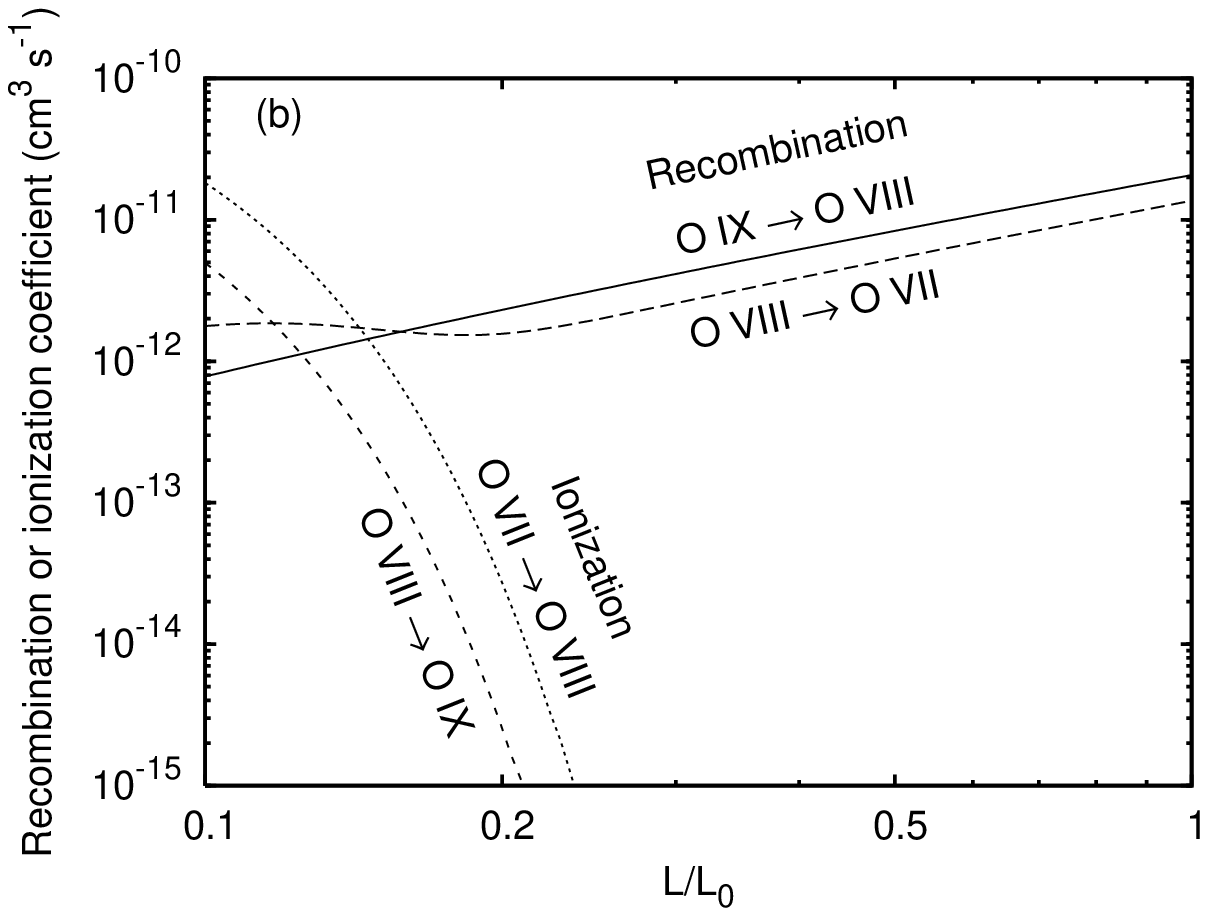}
\caption{(a) Recombination timescales $t_\mathrm{rec}$ for the formation of the labeled ions as a function of Local Bubble size $L$, assuming the
Local Bubble is expanding adiabatically. The recombination timescales and $L$ have been normalized to their present-day values $t_\mathrm{rec,0}$ and $L_0$,
respectively. (b) Recombination and ionization timescales for the labeled processes governing the evolution of the population of \OVIII. In
both figures the curves were calculated assuming a present-day temperature $T_0 = 4.3 \times 10^4$ \K.\label{fig:IonizationTimescale}}
\end{figure}

Figure~\ref{fig:IonizationTimescale}(a) shows $t_\mathrm{rec} / t_\mathrm{rec,0}$ for \OVI, \OVII, and \OVIII\ as functions
of $L/L_0$, assuming a present-day temperature  $T_0 = 4.3 \times 10^4$ \K\ \citep{breitschwerdt01}. Figure~\ref{fig:IonizationTimescale}(b)
shows the recombinations and ionization coefficients for the various process that govern the evolution of the population of \OVIII,
again as functions of $L/L_0$. The ionization coefficients were calculated using data from \citet{arnaud85}.

From Figure~\ref{fig:IonizationTimescale}(a) we can see that the recombination timescale was always shorter at earlier times, when the Local Bubble
was smaller. This is true for all present-day Local Bubble temperatures we investigated. From Figure~\ref{fig:IonizationTimescale}(b), meanwhile, we can see
that when $L \ga 0.15 L_0$, the recombination coefficients governing the evolution of the population of \OVIII\ are larger than the ionization
coefficients (cf.\ in the \citetsq{breitschwerdt94} and \citetsq{breitschwerdt96b} models, the Local Bubble burst out of its natal cloud when
$L \approx 0.2 L_0$). This means that since the Local Bubble burst out, recombination has dominated over ionization in the evolution of the \OVIII\
population. We find this is true in general for $T_0 \la 10^5$ \K. This result, combined with the fact that the timescale
for recombinations was shorter when the Local Bubble was younger, means that the population of \OVIII\ will be depleted more rapidly than in our original
model in which \Ne\ and $T$ are held fixed at their present-day values.

The discussion in the previous paragraph applies only to $T_0 \la 10^5$ \K. For higher present-day temperatures
($T_0 \sim \mbox{few} \times 10^5$ \K), the ionization coefficients that govern the \OVIII\ evolution remain larger than the recombination
coefficients for some time after the Local Bubble has burst out of its natal cloud. However, even for $T_0 = 3 \times 10^5$ \K, recombinations start to
dominate the evolution of \OVIII\ before the Local Bubble reaches half its current size. Furthermore, the much lower recombination timescale at this time
more than compensates for the delay in the start of recombinations -- by the time the Local Bubble reaches its current size, the \OVIII\ population will
be lower than if the ion populations had evolved for the same amount of time with \Ne\ and $T$ held fixed at their present-day values.

The result of the above discussion is that the \OVII\ emission (which in this model is formed by recombinations from \OVIII) will fall below
the observed level on a timescale even shorter than the $\sim$$6 \times 10^5$ \yr\ given by our original model. Hence, as stated above,
$6 \times 10^5$ \yr\ is an upper limit on the time since the Local Bubble burst out of its natal cloud. If the Local Bubble were any older than this, the \OVII\
emission would be below what is observed. 

This age is a factor of 2--7 times smaller than the age of the Local Bubble proposed in recombining models of the Local Bubble
(\citealp{breitschwerdt94}; \citealp{breitschwerdt96b}). It also means that the Local Bubble would have expanded extremely rapidly
after bursting out of its natal cloud: $(100~\pc) / (6 \times 10^5 ~\yr) \approx 160$ \kmps. This is much larger than the adiabatic
sound speed of the ambient medium ($\cs = 15$ \kmps\ when $T = 10^4$ \K), and so the expansion would have been highly supersonic,
resulting in a shock at the edge of the Local Bubble as it expanded, with associated heating and ionization processes. Thus,
the above-described model of an adiabatically cooled, purely recombining Local Bubble is not self-consistent.

It should be noted that strictly the above argument against a recombining model of the Local Bubble only applies for present-day temperatures
$T_0 \la 3 \times 10^5$ \K. However, as noted above, higher present-day temperatures which are observationally acceptable
(see Fig.~\ref{fig:RecombiningModelChisq}) imply a Local Bubble pressure an order of magnitude larger than that of the Local Cloud, in which
case the recombining model loses one of its major selling points. It is also possible that such higher temperatures would still lead
to an implausibly young Local Bubble age. However, to test this would require more detailed calculations of the growth of the Local Bubble and the
ionization evolution therein, which are beyond the scope of this paper.

This is not the first time objections have been raised against a recombining model of the Local Bubble. Using values from \citet{breitschwerdt01},
and taking as a lower limit for the \OVI\ fraction the value for a $5 \times 10^5$ \K\ plasma in equilibrium ($\novi / \nO = 0.037$),
\citet{shelton03} predicted an \OVI\ intensity of $\sim$1900 \lineunit. She adds that this prediction is a lower limit, as the \OVI\
fraction would tend to increase due to recombinations from \OVII, and would not decrease again until most of the oxygen had recombined
to \OI\ and \OII. This predicted intensity is more than twice the $2\sigma$ upper limit on the Local Bubble intensity established with \fuse\
\citep{shelton03}. This, and a number of other discrepancies, led \citeauthor{shelton03} to conclude that recombining models could be
practically eliminated.

\citet{oegerle05} found that a recombining model of the Local Bubble also disagrees with observations from the point of view of \OVI\ absorption.
They found that such a model leads to an \OVI\ column density much larger than that which they had measured with \fuse.

\citet{smith05} considered a large set of recombining plasma models, covering a wide range of electron densities, temperatures, and Local Bubble sizes.
They found that they were unable to match simultaneously the \OVII\ and \OVIII\ intensities they measured from their \chandra\ spectrum of MBM~12,
the upper limit on the \OVI\ intensity \citep{shelton03}, the \OVI\ column density \citep{oegerle05}, and the \rosat\ R12 intensity.
In essence, the R12 intensity requires a minimum density of highly ionized ions; however, the densities of \OVI, \OVII, and \OVIII\ are
limited by the \fuse\ and \chandra\ data. Given this, \citet{smith05} said that a static recombining plasma model (i.e.\ a model in which
the temperature and density governing the recombinations is constant) may be confidently disposed of.

Both \citepossessive{shelton03} and \citepossessive{smith05} objections to a recombining model are based upon the predictions of a static
recombining model. However, \citet{breitschwerdt01} points out that for a proper recombining model of the Local Bubble, the dynamical and thermal
evolution of the Local Bubble must be treated together, self-consistently. While we have not carried out full dynamical modeling of the ionization evolution
in an expanding, cooling Local Bubble, we have shown that our static model gives a Local Bubble age several times smaller than that
in the models of \citet{breitschwerdt94} and \citet{breitschwerdt96b}, and furthermore by considering the recombination timescales when
the Local Bubble was smaller, we have shown that the Local Bubble age given by our static model is an upper limit on the age a dynamical model would give.
To put this another way, if the Local Bubble had burst out of its natal cloud a few million years ago (\citealp{breitschwerdt94}; \citealp{breitschwerdt96b}),
the \OVII\ intensity would be far below what is observed. Our observations therefore add to the evidence against a recombining plasma
model (static or dynamical) of the Local Bubble (at least with a present-day temperature $\la 3 \times 10^5$ \K).


\section{SUMMARY AND CONCLUSIONS}
\label{sec:Summary}

We have analyzed \xmm\ spectra of the interstellar medium, obtained from pointings on and off an absorbing filament at high southern
Galactic latitude. We have fit various models simultaneously to both sets of spectra, and used the difference in the absorbing column
in the two pointing directions to constrain the spectrum of the Local Bubble.

Our main findings are as follows:

1. In the examined direction, The Local Bubble emission is consistent with emission from a thermal plasma in collisional ionization equilibrium with
a temperature $\logTLB = 6.06^{+0.02}_{-0.04}$ and an emission measure $\int \Ne^2 \dl = 0.018$ \emismeas. The temperatures of
the Local Bubble and the Galactic halo components are in good agreement with previous \rosat\ measurements of other directions
(\citealp{snowden98,snowden00}; \citealp{kuntz00}). However, our Local Bubble emission measure is 3--10 times larger than the
\rosat-determined value \citep{snowden98}. This discrepancy is explained by the fact that we use a different plasma emission code
and abundance table from \citet{snowden98}.

2. Our Local Bubble temperature disagrees with the results of \xmm\ observations of the Local Bubble in other directions, which find
$\logTLB \approx 6.2$ (\citealp{freyberg03}; \citealp{freyberg04a}; \citealp{freyberg04b}). If we use a model with this
higher $\TLB$, we find it over-predicts the Galactic halo \OVI\ emission by several orders of magnitude. This therefore
suggests that the Local Bubble is thermally anisotropic. However, it is possible that for some of these other \xmm\ observations
the foreground emission is being contaminated by non-Local Bubble emission from Loop I.

3. Our data are also consistent with a non-equilibrium model in which the plasma is underionized. However, while an overionized recombining
plasma model is observationally acceptable for certain densities and temperatures, it generally gives a very young
age for the Local Bubble: $\la 6 \times 10^5$ \yr. This is several times lower than the Local Bubble age in the models of
\citet{breitschwerdt94} and \citet{breitschwerdt96b}. Such a young age is implausible, as it would require the Local Bubble to have burst
highly supersonically out of its natal cloud, meaning that a purely recombining model is not self-consistent.

As stated in the Introduction, X-ray spectroscopy is essential for distinguishing between models of Local Bubble formation. The \xmm\
observations presented here have added to the evidence against overionized, recombining models of the Local Bubble. Forthcoming 
\suzaku\ data from these same viewing directions will extend our sensitivity to lower energies, enabling us to place constraints
on the carbon and nitrogen line emission in the $\sim$0.3--0.5 \kev\ energy range. This increased spectral information will help us
constrain the other major class of non-equilibrium Local Bubble model, namely an underionized plasma that is in the process of ionizing.

\section*{} 

We would like to thank Dan McCammon, Bart Wakker, Yangsen Yao, and the referee, Joel Bregman, for helpful comments and suggestions.
This work is based on observations obtained with \xmm, an ESA science mission with instruments and contributions directly funded by ESA Member States and NASA.
This work was funded by NASA grant NNG04GD78G (awarded through the Long Term Space Astrophysics program) and NASA grants NNG04GB68G
and NNG04GB08G (awarded through the \xmm\ Guest Investigator Program).


\appendix
\section{MEASURING \TLB\ FROM FREYBERG ET AL.'S (2004) BARNARD~68 DATA}
\label{app:Barnard68}

\citet{freyberg04b} do not quote a value for \TLB\ measured from their \xmm\ observation
of the Bok globule Barnard~68. Here we describe how a Local Bubble temperature may be inferred from the data they do present.

Barnard~68 casts a deep shadow in the soft X-ray background. However, the shadowing is not complete at \OVIII\ energies, implying at least
some of the \OVIII\ emission originates in front of Barnard~68. If one attributes this emission to the Local Bubble, then from
\citepossessive{freyberg04b} Figure~2 (which gives the ratio of the on-cloud to off-cloud intensity as a function of energy) one
can estimate the \OVII:\OVIII\ intensity ratio for the Local Bubble emission, and hence infer a temperature.

The ratio \Roviii\ of on-cloud to off-cloud intensity for \OVIII\ emission is
\begin{equation}
	\Roviii = \frac{\Loviii + \mathrm{e}^{-\taueon} \Boviii}{\Loviii + \mathrm{e}^{-\taueoff} \Boviii},
\label{eq:R8}
\end{equation}
where \Loviii\ and \Boviii\ are the Local Bubble and background \OVIII\ intensities\footnote{Note that \Boviii\ will in fact consist of
\OVIII\ emission from the Galactic halo, and also continuum emission from the extragalactic background.}, and \taueon\ and \taueoff\
are the on- and off-cloud optical depths at \OVIII\ energies. There is a similar equation for \Rovii, the corresponding
ratio for \OVII\ emission. The on- and off-cloud column densities are $\sim$$10^{23}$ and $\sim$$10^{20}$ \pcmsq\ \citep{freyberg04b}.
We calculate absorption cross-sections per hydrogen atom using the \citetsq{balucinska92} cross-sections (except for He; \citealp{yan98})
with the \citetsq{wilms00} interstellar abundances. The values at 0.57 \kev\ (\OVII) and 0.654 \kev\ (\OVIII) are $6.96 \times 10^{-22}$ and
$4.74 \times 10^{-22}$ \cmsq, respectively. Given the large
on-cloud optical depths, we can ignore the term involving \taueon\ in equation~(\ref{eq:R8}), and the corresponding term involving
\tauson\ in the equation for \Rovii. Hence, by combining equation~(\ref{eq:R8}) with its \Rovii\ analogue we find that
\begin{equation}
	\frac{\Loviii}{\Lovii} = \frac{\Boviii \Roviii \mathrm{e}^{-\taueoff}}{\Bovii \Rovii \mathrm{e}^{-\tausoff}}
				 \left( \frac{1 - \Rovii}{1- \Roviii} \right).
\label{eq:L8overL7}
\end{equation}
Figure~2 in \citet{freyberg04b} plots the on-off intensity ratio every 50 \ev. Thus, for the purposes of this estimate, we read off
\Rovii\ and \Roviii\ at 0.55 and 0.65 \kev, respectively (yielding $\Rovii = 0.71$ and $\Roviii = 0.51$), and we take the \OVII\ and
\OVIII\ ``bands'' to be 0.525--0.575 and 0.625--0.675 \kev, respectively. If we assume the Galactic halo and extragalactic background
are isotropic, we can use our ``standard'' model results in Table~\ref{tab:FitResults} to estimate that $\Boviii / \Bovii = 0.455$.
We therefore obtain $\Loviii / \Lovii = 0.20$. By using XSPEC to calculate APEC model fluxes in the above energy bands, we find that this
ratio corresponds to $\logTLB = 6.24$ (we obtain the same value if we use \mekal\ model fluxes, instead of APEC). By using a full
plasma emission model to calculate flux ratios, as opposed to simply using the \OVII\ and
\OVIII\ line emissivities, we also take into account the contribution of continuum emission to the two energy bands being considered.

\begin{figure}
\centering
\includegraphics[width=0.45\linewidth]{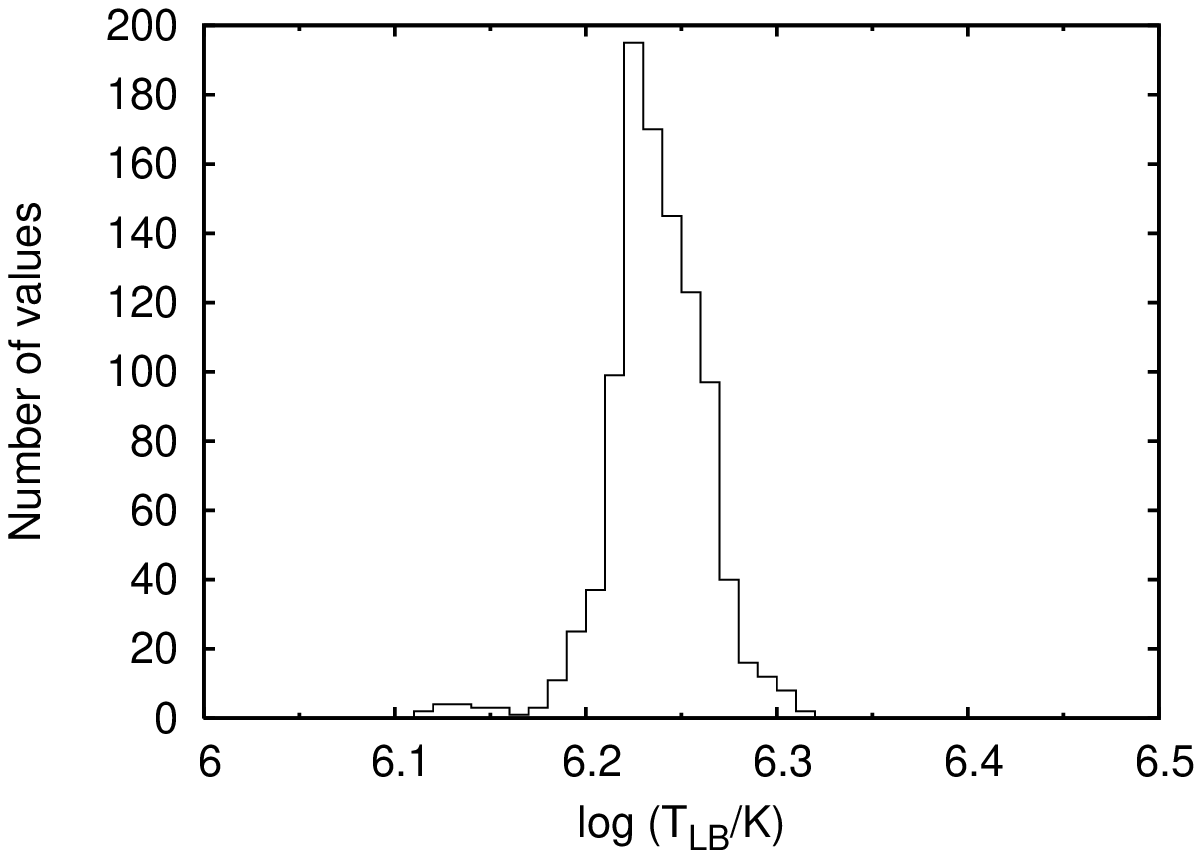}
\caption{Histogram of 1000 randomly generated values of \logTLB, based upon the \xmm\ observation of Barnard 68 \citep{freyberg04b}.
The temperatures were calculated from 1000 values $\Loviii / \Lovii$, which in turn were calculated using equation~(\ref{eq:L8overL7}) assuming
20\%\ errors on $\Boviii / \Bovii$, \taueon, and \tauson.\label{fig:Barnard68Temperature}}
\end{figure}

This estimate of \TLB\ assumes that the value of $\Boviii / \Bovii$ measured from our \xmm\ spectra can
be applied to this pointing direction. We have also assumed a single absorption cross-section for each energy band, when in fact the
absorption cross-section will vary across each band. To
estimate how much of an effect these assumptions have on our result, we used a Monte Carlo method similar to that described in
\S\ref{subsec:Halo} to calculate 1000 values of \TLB\ assuming 20\%\ errors on $\Boviii / \Bovii$, \taueon, and \tauson. Visual inspection
of the resulting histogram of values indicates that this estimate of \TLB\ is accurate to within 0.1 dex (see Fig.~\ref{fig:Barnard68Temperature}).

\end{document}